\documentclass[aps,prd,notitlepage,nofootinbib,superscriptaddress]{revtex4}
\usepackage{amssymb}
\usepackage{color}
\usepackage{amsmath}
\usepackage{verbatim}
\usepackage{appendix}
\usepackage[usenames,dvipsnames]{xcolor}

\newcommand{\fppx}{\hat{F}^{x,ab}_{\varphi_a\varphi_b}}

\newcommand{\cfppx}{\hat{{\cal F}}^{x,ab}_{\varphi_a\varphi_b}}
\newcommand{\cfppy}{\hat{{\cal F}}^{y,ab}_{\varphi_a\varphi_b}}
\newcommand{\gpp}{G^{xy}_{\varphi_a\varphi_b, cb}}

\newcommand{\pa}{\pi_\Lambda^{ij}}
\newcommand{\vx}{\vec{x}}
\newcommand{\vy}{\vec{y}}
\newcommand{\vz}{\vec{z}}
\newcommand{\faa}{{\hat{F}_{\Lambda\Lambda}}}
\newcommand{\cfaa}{{\hat{{\cal F}}_{\Lambda\Lambda}}}
\newcommand{\dxy}{\delta^3(\vec{x}-\vec{y})}
\newcommand{\fan}{{\hat{F}_{\Lambda N}}}
\newcommand{\cfan}{{\hat{{\cal F}}_{\Lambda N}}}
\newcommand{\vxx}{\vec{x}'}
\newcommand{\ch}{{\cal H}_i}
\newcommand{\mom}{\tilde{{\cal H}}_i}
\newcommand{\tc}{\tilde{\chi}}
\newcommand{\bN}{\bar{N}}
\newcommand{\bh}{\bar{h}}
\newcommand{\bK}{\bar{K}}
\newcommand{\bp}{\bar{\pi}}
\newcommand{\tbp}{\bar{\tilde{\pi}}}
\newcommand{\cn}{C_{\pi_N}}
\newcommand{\ci}{C_{\pi_i}}
\newcommand{\ca}{C_{\pi_\Lambda}}
\newcommand{\cph}{C_{\pi_h}}

\begin{document}

\title{Canonical invariance of spatially covariant scalar-tensor theory}

\author{Rio Saitou}
\email{riosaitou@hust.edu.cn}
\affiliation{School of Physics, Huazhong University of Science and Technology, Wuhan 430074, China}


\begin{abstract}
We investigate the invariant canonical transformation of a spatially covariant scalar-tensor theory of gravity, called the XG theory.
Under the invariant canonical transformation, the forms of the action or the Hamiltonian and the primary constraints are preserved,
but the action or the Hamiltonian is not invariant.
We derive the Hamiltonian in a non perturbative manner and perfrom the Hamiltonian analysis for full theory.
We confirm that the theory has at most 3 degrees of freedom as long as the theory has the spatial diffeomorphism symmetry.
Then, we derive the invariant canonical transformation by using the infinitesimal transformation.
The invariant metric transformation of the XG theory contains a vector product as well as the disformal transformation.
The vector product and the disformal factor can depend on the higher order derivatives of the scalar field and the metric.
Furthermore, we discover additional canonical transformation which leaves the theory invariant.
Using the invariant transformation, we study the relation between the Horndeski theory and the GLPV theory,
and find that we can not obtain general GLPV theory from the Horndeski theory through the invariant canonical transformation we found.
\end{abstract}

\maketitle

\section{Introduction}\label{sec1}


It is well known that general relativity is the most successful theory of gravity which contains two tensor degrees of freedom.
We do not need, however, to stick with the gravitational theory containing two tensor modes if we consider more
general low energy effective field theories of gravity or try to explain the mysteries of the universe such as the cosmic acceleration.
As an extension to Einstein's general relativity,
we can consider the scalar-tensor theory of gravity in which we add a scalar degree of freedom interacting with the tensor modes.
The idea of adding a scalar mode to gravitational theory was first proposed in Brans-Dicke theory \cite{Brans:1961sx} in the early 1960's, and thereafter, numerous scalar-tensor theories which add one or more scalar modes were devised [2--15].

Among those scalar-tensor theories, the Horndeski theory \cite{Horndeski:1974wa,Deffayet:2009wt,Kobayashi:2011nu} was regarded as the most general stable
scalar-tensor theory containing one scalar and two tensor modes. Although the theory has more than two time derivatives in the action,
the equations of motion remain to be second order differential equations. In general, if we have higher than second order time derivatives
in the equations of motion, there appears extra degrees of freedom, the so called Ostrogradski ghosts \cite{Woodard:2006nt}.
The ghost makes the Hamiltonian unbound from below, so that the system becomes unstable. The equations of motion of the  Horndeski theory,
however, are limited up to second order, and thus the Horndeski theory can avoid the ghost.
In the frame of the Horndeski theory,
various cosmological models have been suggested, some of which have characteristic mechanisms to drive the accelerating expansion of the universe \cite{Deffayet:2010qz, Kobayashi:2010cm}.

Although the Horndeski theory was regarded as the most general stable scalar-tensor theory
containing one scalar and two tensor modes, Gleyzes et al. recently suggested a new scalar-tensor theory beyond the Horndeski theory, called the GLPV theory \cite{Gleyzes:2014dya},
in which the equations of motion include third order time derivatives. Nevertheless, several Hamiltonian analyses showed that
there appears no Ostrogradski ghost at least for the restricted classes of the GLPV theory \cite{Lin:2014jga, Deffayet:2015qwa, Langlois:2015skt}.
After the GLPV theory has came out, a lot of efforts has been made to search for stable scalar-tensor theories beyond the Horndeski theory. The spatially covariant scalar-tensor theory \cite{Gao:2014soa}, which we call the XG theory, is one of the results. The XG theory is an effective field theory which has the symmetry 
of the spatial diffeomorphism only, hence the theory consists of the ADM variables \cite{Arnowitt:1962hi} and the time coordinate $t$ explicitly. Therefore, the XG theory includes Ho\v{r}ava gravity \cite{Horava:2009uw} and its extensions \cite{Blas:2010hb, Colombo:2014lta} although they were constructed in different context.
The Horndeski and the GLPV theories have the gauge symmetry of the spacetime diffeomorphism. However, if we take a preferred foliation of spacetime
by identifying the timelike scalar field $\phi$ of the theories to be the time coordinate $t$, the theories become subclasses of the XG theory. Such identification could be interpreted as a (partial) gauge fixing, and we call it the unitary gauge. Conversely, if we can restore the symmetry of the spacetime diffeomorphism by introducing the scalar field $\phi$ in the XG theory,
then the XG theory can be expressed as a general covariant theory. Thus, the XG theory is a more general theory including both the Horndeski theory and the GLPV theory if we can restore the gauge symmetry by the scalar field $\phi$. We will review the XG theory in more detail in the next section.


Until now, many efforts have been devoted to construct new stable scalar-tensor theories beyond the Horndeski theory. In parallel with it,
the connection between different theories of gravity has been also investigated through the
transformation of the spacetime metric \cite{Bekenstein:1992pj, Nojiri:2003rz, Gleyzes:2014qga, Bettoni:2013diz}.
Practically, it is really beneficial to consider how a theory of gravity transforms to another by the metric transformation.
Transforming the action of one theory to a well-known action via the metric transformation, we can sometimes carry out calculations easily,
or we can intuitively find out the degrees of freedom the theory has\cite{Nojiri:2003rz}.
Moreover, through the metric transformation, we can easily understand the screening mechanism of the scalar degree
of freedom \cite{Chameleon, Kobayashi:2014ida}, the formation of stars with the scalar degree of freedom \cite{Koyama:2015oma, Saito:2015fza}, and so on.
On the other hand, to search the metric transformation which leaves the action of the theory invariant means
to search the symmetry of the theory, and it leads to deeper understanding of the theory.
Note that in this article, we regard the theory invariant
\textit{if the action or the Hamiltonian and the primary constraints of the theory are 
transformed to the same family of the theory} under the transformation.
The action of the Horndeski theory is invariant under the disformal transformation \cite{Bettoni:2013diz},
\begin{equation}
\label{disf2}
g_{\mu\nu}\ \rightarrow \ {\cal A[\phi]}g_{\mu\nu} + {\cal B}[\phi]\nabla_\mu\phi\nabla_\nu\phi \ ,
\end{equation}
where the coefficients ${\cal A}$ and ${\cal B}$ are referred to as the conformal factor and the disformal factor respectively.
The GLPV theory is invariant under the same kind of disformal transformation,
but the disformal factor depends on the derivative of the scalar field \cite{Gleyzes:2014qga},
\begin{equation}
\label{disf}
g_{\mu\nu}\ \rightarrow \ {\cal A[\phi]}g_{\mu\nu} + {\cal B}[\phi, X]\nabla_\mu\phi\nabla_\nu\phi, \quad X\equiv -g^{\mu\nu}\nabla_\mu\phi\nabla_\nu\phi \ .
\end{equation}
Interestingly, a particular class of the GLPV theory can be obtained from the Horndeski theory through the
disformal transformation (\ref{disf}) \cite{Gleyzes:2014qga, Crisostomi:2016tcp, Crisostomi:2016czh}.

Having in mind these features of the metric transformation, the following questions naturally arise: Are there more general transformations
other than the disformal transformation which leave scalar-tensor theories invariant or connect different theories?
Furthermore, a little ambitiously, are there transformations which transform the Horndeski theory to arbitrary classes of GLPV theory?
One of the main purposes of this article is to find such general transformations.

When searching for general transformations, we should note that if we can take the unitary gauge $\phi=t$,
all the disformal transformations we mentioned above reduce to the point transformations of the ADM variables.
It has been well known that the point transformations belong to more general transformations, the canonical transformations.
Hence, in this article, we aim to derive the canonical transformation which leaves the XG theory invariant,
and to reveal the relation between the GLPV and the Horndeski theories in the unitary gauge using the invariant canonical transformation of the XG theory.

In order to investigate the structure of the XG theory and the relation between the Horndeski and the GLPV theories,
we take the following steps. Firstly, we derive the non-perturbative Hamiltonian and perform the Hamiltonian analysis for the XG theory.
Until now, the Hamiltonian analysis of the XG theory was done only in a perturbative way \cite{Gao:2014fra}.
Secondly, we search the canonical transformation which leaves the XG theory invariant.
Finally, we investigate the relation between the Horndeski and the GLPV theories in the unitary gauge focusing on
whether we can obtain general GLPV theory from the Horndeski theory using the invariant transformation of the XG theory.

The disformal transformation (\ref{disf2}) is the most general metric transformation which leaves the Horndeski theory invariant.
Under the disformal transformation (\ref{disf2}), the equations of motion remain to be second order differential equations.
However, we know that the equations of motion of several stable scalar-tensor theories beyond the Horndeski theory are differential equations with higher than second order derivatives,
and hence we can not use the strategy keeping the equations of motion at second order to find the invariant transformation.
We consider the infinitesimal transformation first because it is easier to work on. If a certain infinitesimal transformation leaves the theory invariant,
then we repeat the infinitesimal transformation to get the finite transformation which leaves the theory invariant.
We can not obtain, however, any discrete transformation from the infinitesimal transformation.

This paper is organized as follows. We briefly review the XG thoery in Section \ref{sec2}. The non-perturbative Hamiltonian analysis of the XG theory was performed in Section \ref{sec3}.
We derive the infinitesimal transformation which leaves the XG theory invariant in Section \ref{sec4}, and promote it to the finite transformation if possible.
Then, we investigate the relation between the Horndeski and the GLPV theories in the unitary gauge with the infinitesimal transformation in Section \ref{sec5}.  Section \ref{sec6} is devoted to the summary of this article. In  Appendix A, we complement the result in Section \ref{sec3} by performing the Hamiltonian analysis for two special cases of the XG theory. In Appendix B and C, we give  detailed calculations for the quantities we use in Sections \ref{sec4} and \ref{sec5} respectively.

\section{The XG theory}\label{sec2}

In this section, we briefly review the construction of the XG theory \cite{Gao:2014soa}.
The XG theory respects the 3 dimensional spatial diffeomorphism, and thus
the ingredients of the XG theory consist of the ADM variables
\begin{equation}
ds^2 = -N^2dt^2 + h_{ij}(N^idt+dx^i)(N^jdt+dx^j) \ ,
\end{equation}
where $N(x)$ is the lapse function, $N^i(x)$ is the shift function, and the Latin index represents the spatial coordinates.
The  ADM variables transform under the three spatial infinitesimal diffeomorphism, $\delta x^i = \xi^i(\vx,t)$, as
\begin{align}
\label{va}
    \delta N&= \xi^iD_iN  \ , \nonumber \\
    \delta N^i&= \eta^i + {\frak L}_\xi N^i,\quad \eta^i\equiv\dot{\xi}^i \ , \nonumber \\
    \delta h_{ij}&= {\frak L}_\xi h_{ij} \ ,
\end{align}
where $D_i$ is the covariant derivative on the spatial hypersurface compatible with the spatial metric $h_{ij}$, the dot represents the time derivative, and ${\frak L}_\xi$ is the Lie derivative along the 3-vector $\xi^i$.
Using the ADM variables, we define the extrinsic curvature
\begin{equation}
K_{ij} = \frac{1}{2N}\left(\dot{h}_{ij} - D_iN_j - D_jN_i\right) \ .
\end{equation}
To preserve the spatial diffeomorphism, we restrict the building blocks of the theory as follows:
\begin{equation}
\label{bb}
\sqrt{h}d^3x,\  t,\ N, \ h_{ij}, \ K_{ij},\ R_{ij},\ D_i \ ,
\end{equation}
where $h$ is the determinant of $h_{ij}$ and $R_{ij}$ is the Ricci curvature with respect to $h_{ij}$.
We do not include the Riemann tensor since we can represent it by the combination of $R_{ij}$ and $h_{ij}$ in 3 dimensions.
Then, the action of the XG theory is given as
\begin{equation}
\label{S}
S = \int dtd^3xN\sqrt{h}\,F[N,\, h_{ij},\, K_{ij},\, R_{ij},\,D_i,\, t] \ ,
\end{equation}
where $F$ is a general scalar functional of its arguments.
Note that this action is \textit{not} the most general spatially covariant action consisted of the ADM variables since we do not include the time derivative of the building blocks. To preserve the symmetry under the spatial diffeomorphism, for example, we can always include a combination
\begin{equation}
\label{ }
N_{\bot} = \dot{N} - N^iD_iN
\end{equation}
into the building blocks since it behaves as a scalar under the spatial diffeomorphism.
We do not usually consider the time derivative of the lapse function since if it exists, there often appears a ghost particle which makes the theory unstable.
But, the ghost does not always appear, and thus we may be able to include the time derivative of the
lapse function while keeping the theory stable \cite{Domenech:2015tca}.
We should need, however, some fine tunings of the kinetic term of the theory to exclude the ghost. Therefore, we may regard the XG theory as \textit{a stable spatially covariant scalar tensor theory without a fine tuning to its kinetic term}\footnote{We can sometimes avoid the instability without the fine tuning even if there exists the ghost. When the mass of the ghost is heavier than the cutoff energy scale of the low energy effective theory, we can safely ignore the ghost in the low energy regime as in the ghost condensation \cite{ArkaniHamed:2003uy}.}.



The XG theory is a nonrelativistic theory because it explicitly breaks the spacetime diffeomorphism.
If we regard the XG theory as a gauge fixed form of a general relativistic theory, however, we can find the relativistic form of the XG theory by restoring the broken gauge symmetry.
In the XG theory, the spacetime diffeomorphism is broken by choosing a preferred spacetime foliation which fixes the time coordinate.
Thus, we can restore the full spacetime diffeomorphism if we introduce a gauge degree of freedom for the choice of time. To do this,
we introduce the St\"uckelberg field $\phi$ as the role of time. 
We regard the constant time hypersurface as the hypersurface on which $\phi$ has a constant value, with the unit normal vector given as
$n_\mu= -\nabla_\mu\phi/\sqrt{X}$ where $\nabla_\mu$ is the covariant derivative compatible with the spacetime metric $g_{\mu\nu}$.
Then, all of spatially covariant quantities can be rewritten in the general covariant form with the spacetime metric $g_{\mu\nu}$ and the scalar field $\phi$:
\begin{align}
\label{}
    t&\rightarrow \phi \ , \nonumber \\
    N&\rightarrow \frac{1}{\sqrt{X}} \ , \nonumber   \\
    h_{ij}&\rightarrow h_{\mu\nu} = g_{\mu\nu} + n_\mu n_\nu = g_{\mu\nu}+\frac{\nabla_\mu\phi\nabla_\nu\phi}{X}\ ,  \nonumber \\
    K_{ij}& \rightarrow K_{\mu\nu}= h^{\rho}_\mu\nabla_{\rho}n_{\nu}= -\frac{1}{\sqrt{X}}
    \left( \nabla_\mu\nabla_\nu\phi - \nabla_{(\mu}\phi\nabla_{\nu)}{\rm ln}X
    - \frac{1}{2X}\nabla_\mu\phi\nabla_\nu\phi\nabla^\rho\phi\nabla_\rho{\rm ln}X \right) \
\end{align}
and so on. We can replace $R_{ij}$ to its general covariant form by using the Gauss-Codazzi relation.
As a consequence, we will obtain the covariant form of the action for the XG theory as
\begin{equation}
\label{SXG}
S_{XG} = \int d^4x\sqrt{-g}{\cal L}[\phi, \nabla_\mu\phi, \nabla_\mu\nabla_\nu\phi, \cdots, g_{\mu\nu}, R_{\mu\nu\rho\sigma}, \nabla_{\mu_1}R_{\mu\nu\rho\sigma}, \cdots] \ ,
\end{equation}
where `$\cdots$' represents higher order derivative terms of $\phi$ and $R_{\mu\nu\rho\sigma}$ is the spacetime Riemann tensor.
Assuming the gradient of the scalar field is timelike, $g^{\mu\nu}\nabla_\mu\phi\nabla_\nu\phi<0$, we can take the unitary gauge in which we choose $\phi=t$, and we can find the spatially covariant action (\ref{S}). Note that the general covariant form of action (\ref{SXG}) is tuned so that it will not include any quantities other than the building blocks (\ref{bb}) when we take the unitary gauge.

\section{Non-perturbative Hamiltonian analysis of the XG theory}\label{sec3}


%
The Hamiltonian of XG theory was derived only in a perturbative way and the same is true for the Hamiltonian analysis \cite{Gao:2014fra}.
The Hamiltonian constructed in the perturbative way is valid only when the amplitudes of fluctuations of the fields are so small that the interactions are suppressed.
Therefore, we need to derive the Hamiltonian in a non-perturbative way.
In this section, we derive the non-perturbative Hamiltonian with an auxiliary tensor field,
and elucidate the structure of the constraints and the number of physical degrees of freedom by the Hamiltonian analysis.

In general, the kinetic term in the XG theory (\ref{S}) is not quadratic, so we can not define the conjugate momenta in the usual way
and we are unable to derive the non-perturbative Hamiltonian from the action.
Then, to make the kinetic term to be lower than quadratic and define the conjugate momenta, we introduce an auxiliary field $\Lambda_{ij}$ for $K_{ij}$ and rewrite the action as
\begin{equation}
\label{SA}
S_\Lambda = \int dtd^3xN\sqrt{h}\, \left\{ F_\Lambda^{ij}(K_{ij} - \Lambda_{ij}) + F[N,\, h_{ij},\, \Lambda_{ij},\, R_{ij},\,D_i,\, t] \right\} \ ,
\end{equation}
where we define the first order functional derivative of $F$ as
\begin{equation}
\label{Fq}
F_{\varphi_a}^a\equiv \frac{1}{N\sqrt{h}}\int d^3x'\frac{\delta (N\sqrt{h}F)(\vec{x}', t)}{\delta \varphi_a(\vx, t)},
\quad \varphi_a=(N, N^i, h_{ij}, \Lambda_{ij}) \ ,
\end{equation}
and the index $a$ on $F$ corresponds to the index of the component field of $\varphi_a$. By construction,
the auxiliary field $\Lambda_{ij}$ is a tensor field under the spatial diffeomorphism. For later convenience,
we introduce the functional derivative of $F_{\varphi_a}^a$ 
\begin{align}
\label{}
   \fppx[\delta^3(\vx-\vy)] \equiv \sqrt{h}\frac{\delta F_{\varphi_a}^a(\vx, t)}{\delta \varphi_b(\vy,t)} \ .
\end{align}
In general, $\fppx$ is a derivative operator with respect to the spatial coordinates $\vx$.
We define another derivative operator $\cfppx$ using $\fppx$ as
\begin{align}
\label{cf}
   \int d^3x\, \cfppx[T(\vx, t)]\delta^3(\vx-\vy)  &\equiv \int d^3x\,T(\vx,t)\fppx[\delta^3(\vx-\vy)] \nonumber \\
    &=\cfppy[T(\vy,t)]   \ ,
\end{align}
where $T(\vx,t)$ is a generic tensor field.
$\cfppx$ is obtained from $\fppx$ performing the integration by parts. 
We can write the second line of (\ref{cf}) as
\begin{equation}
\label{ }
\cfppy[T(\vy,t)] = \int d^3x T(\vx, t)\cfppy[\delta^3(\vy-\vx)] \ , \nonumber
\end{equation}
thus, we obtain the formal relation
\begin{equation}
\label{rel}
 \fppx[\delta^3(\vx-\vy)] = \cfppy[\delta^3(\vx-\vy)] \ .
\end{equation}
Note that each side of the above relation takes differentiation with respect to different variables $\vx$ and $\vy$.
Another useful relation is obtained by integrating (\ref{cf}) with respect to $\vy$
\begin{align}
\label{}
    \int d^3yd^3x\,U(\vy, t)T(\vx,t)\fppx[\delta^3(\vx-\vy)]
             &= \int d^3x\,T(\vx,t)\fppx\left[\int d^3y\, U(\vy, t)\delta^3(\vx-\vy)\right]  \nonumber \\
             &= \int d^3x\, T(\vx, t)\fppx[U(\vx,t)]   \nonumber \\
             &=\int d^3y\,U(\vy,t)\cfppy[T(\vy,t)] \ , \nonumber
\end{align}
where $U(\vx, t)$ is another generic tensor field.
We are free to change the coordinates of the integration since the interval of the integration covers the whole space, so
\begin{equation}
\label{ }
\int d^3x\, T(\vx, t)\fppx[U(\vx,t)]  = \int d^3x\,U(\vx,t)\cfppx[T(\vx,t)] \ .
\end{equation}
Furthermore, we define the Green's function of the derivative operator $\fppx$, if it exists, as
\begin{align}
\label{ }
\fppx [\gpp] &= \mathbb{I}^a_c\delta^3(\vx-\vy) \ , \nonumber \\
\int d^3x \fppx[\gpp]T(\vx, t) &= \int d^3x \cfppx[T(\vx, t)]\gpp = \mathbb{I}^a_c T(\vy, t) \ ,
\end{align}
where $\mathbb{I}^a_c$ is the identity operator. Using them, we immediately obtain the following relation
\begin{align}
\label{FG}
    \int d^3x \fppx[\gpp]G^{zx}_{\varphi_a\varphi_b, ad} &= \int d^3x \gpp \cfppx[G^{zx}_{\varphi_a\varphi_b, ad}]
       = G^{zy}_{\varphi_a\varphi_b, cd} \  , \nonumber  \\
    \cfppx[G^{yx}_{\varphi_a\varphi_b, ad}] &= \mathbb{I}^b_d\delta^3(\vx-\vy)  \ .
\end{align}

With these notations, the variation of the action (\ref{SA}) with respect to $\Lambda_{ij}$ gives the equation
\begin{equation}
\label{ }
\frac{\delta S_\Lambda}{\delta \Lambda_{ij}(\vx,t)} = \hat{{\cal F}}^{x, ij,kl}_{\Lambda\Lambda}[N(K_{kl}-\Lambda_{kl})]
    -N\sqrt{h}F_\Lambda^{ij}+N\sqrt{h}F_\Lambda^{ij} =0\ ,
\end{equation}
and if $\hat{{\cal F}}^{x, ij,kl}_{\Lambda\Lambda}\neq 0$, we obtain a solution
\begin{equation}
\label{ }
\Lambda_{ij} = K_{ij} \ .
\end{equation}
Substituting this solution back to the action (\ref{SA}), we recover the original action (\ref{S}).
The condition $\hat{{\cal F}}^{x, ij,kl}_{\Lambda\Lambda} \neq 0$ should be satisfied for any
dynamical system, otherwise it has no kinetic term. Thus, we can conclude that the rewritten action (\ref{SA}) is equivalent to the original action at least classically.
By using the action (\ref{SA}), we get the conjugate momenta
\begin{equation}
\label{ }
\pi_N \equiv \frac{\delta S_\Lambda}{\delta \dot{N}}=0,\quad \pi_i \equiv \frac{\delta S_\Lambda}{\delta \dot{N^i}}=0,
\quad \pa\equiv \frac{\delta S_\Lambda}{\delta \dot{\Lambda}_{ij}}=0,\quad
\pi_h^{ij}\equiv \frac{\delta S_\Lambda}{\delta \dot{h}_{ij}}=\frac{\sqrt{h}}{2}F_\Lambda^{ij}\ .
\end{equation}
The Hamiltonian is
\begin{align}
\label{H}
H &\equiv \int d^3x \left[ \pi_h^{ij}\dot{h}_{ij} - {\cal L} \right] = \int d^3x \left[ N^i{\cal H}_i + {\cal H}_{\bot}\right] \ , \nonumber \\
{\cal H}_i &\equiv -2\sqrt{h}D_j\left(\frac{{\pi_h}_i^j}{\sqrt{h}}\right) \ , \nonumber  \\
{\cal H}_{\bot}&\equiv N\sqrt{h}(F_\Lambda^{ij}\Lambda_{ij} - F)
\end{align}
where $\pi^j_i = h_{ik}\pi^{jk}$. Note that the spatial indices are raised and lowered by the spatial metric $h_{ij}$.
For example, the trace of a tensor is $\pi_h\equiv \pi_h^{ij}h_{ij}$. The XG theory has the gauge symmetry under the spatial diffeomorphism, hence the system has primary constraints
\begin{equation}
\label{pri}
\pi_N \approx0,\quad \pi_i \approx0,
\quad \pa\approx0,\quad
\tilde{\pi}_h^{ij}\equiv \pi^{ij}_h - \frac{\sqrt{h}}{2}F_\Lambda^{ij} \approx 0 \ ,
\end{equation}
where `$\approx$' represents the weak equality. We define the total Hamiltonian with the Lagrange multipliers as
\begin{equation}
\label{Ht}
H_{tot} \equiv H + \int d^3x \left( \lambda_N\pi_N + \lambda^i\pi_i +\lambda_{\Lambda ij}\pi_\Lambda^{ij}
+ \lambda_{hij}\tilde{\pi}_h^{ij}\right) \ .
\end{equation}
Before starting the Hamiltonian analysis, we define the Poisson bracket and the inner product for generic tensor fields $X(\vx)$ and $Y(\vx)$ as
\begin{align}
\label{ }
\{X(\vx),\,Y(\vy)\}_P &\equiv \sum_{a}\int d^3z\left(\frac{\delta X(\vx)}{\delta \varphi_a(\vz)}\frac{\delta Y(\vy)}{\delta \pi_{a}(\vz)} - \frac{\delta X(\vx)}{\delta \pi_{a}(\vz)}\frac{\delta Y(\vy)}{\delta \varphi_a(\vz)}\right) \ , \\
 \left< X(\vx)\,,\, Y(\vx)\right> &\equiv \int d^3x X(\vx)Y(\vx),
\end{align}
where $\pi_{a}$ are the conjugate momenta of $\varphi_a$. The quantities in the Poisson bracket and the inner product are given at the same time.
Using these definitions as well as Eqs. (\ref{Fq})-(\ref{FG}), we are ready to perform the Hamiltonian analysis.
We find that the Poisson brackets of the following primary constraints are generally nonzero,
\begin{align}
\label{}
    \{ \pi_N(\vx),\, \tilde{\pi}_h^{ij}(\vy)\}_P&= \frac{1}{2}\hat{F}^{y,ij}_{\Lambda N}[\delta^3(\vx-\vy)] \ ,  \nonumber \\
    \{ \pi_\Lambda^{ij}(\vx),\, \tilde{\pi}_h^{kl}(\vy)\}_P&= \frac{1}{2}\hat{F}_{\Lambda\Lambda}^{y,kl,ij}[\dxy] \ , \nonumber \\
    \{ \tilde{\pi}_h^{ij}(\vx),\, \tilde{\pi}_h^{kl}(\vy)\}_P&= \frac{1}{2}\left( \hat{F}_{\Lambda h}^{y,kl,ij}[\dxy]-
    \hat{F}_{\Lambda h}^{x, ij,kl}[\dxy]\right) +\frac{\sqrt{h}}{4}\left( h^{ij}F_\Lambda^{kl}- h^{kl}F_\Lambda^{ij}\right)\dxy
    \nonumber \\
    &\equiv \hat{\Pi}^{ij,kl}[\dxy] \ ,
\end{align}
and the other Poisson brackets of the primary constraints are zero.
We assume hereafter that the Green's function of $\faa^{x, ij,kl}$, i.e. $G_{\Lambda\Lambda, ij,mn}^{xy}$, always exists.  Even if this assumption is not true, the final result of the Hamiltonian analysis will not change substantially (see Appendix \ref{app1} on this point). If $G_{\Lambda\Lambda, ij,mn}^{xy}$ exists, we can make a linear combination of $\pi_N$ and $\pi_\Lambda^{ij}$
\begin{equation}
\label{pineq1}
\tilde{\pi}_N\equiv \pi_N - \cfan^{x,mn}\left[ \left<G^{xx'}_{\Lambda\Lambda, mn,ij}\,,\, \pi_\Lambda^{ij}(\vec{x}')\right>\right] \ .
\end{equation}
Taking notice of the following relations
\begin{align}
\label{}
    \left\{ \left< 2G^{x'x}_{\Lambda\Lambda, mn,ij}\,,\,\pi_\Lambda^{ij}(\vec{x}')\right>,\,\tilde{\pi}_h^{kl}(\vy)\right\}_P&\approx
    \int d^3x' G^{x'x}_{\Lambda\Lambda,mn,ij}\faa^{y,kl,ij}[\delta^3(\vy-\vec{x}')] \nonumber \\
    &= \faa^{y,kl,ij}\left[\int d^3x' G^{x'x}_{\Lambda\Lambda,mn,ij}\delta^3(\vy-\vec{x}')\right]
    =\mathbb{I}^{kl}_{mn}\dxy \ ,  \\
    \label{FGI}
   \left \{ \tilde{\pi}_\Lambda^{ij}(\vx)\,,\,\left< 2G^{yx'}_{\Lambda\Lambda, kl,mn}\,,\,\tilde{\pi}_h^{kl}(\vec{x}')\right>\right\}_P&\approx
    \int d^3x' G^{yx'}_{\Lambda\Lambda,kl,mn}\faa^{x',kl,ij}[\delta^3(\vec{x}'-\vec{x})]  \nonumber \\
    &= \int d^3x' \cfaa^{x',kl,ij}[G^{yx'}_{\Lambda\Lambda,kl,mn}]\delta^3(\vec{x}'-\vec{x}) = \mathbb{I}^{ij}_{mn}\dxy \ ,
\end{align}
where we use the relation (\ref{FG}) for the last equality in Eq. (\ref{FGI}), we find that
the Poisson brackets of $\tilde{\pi}_N$ with all the primary constraints become zero,
\begin{align}
\label{las}
    &\{\tilde{\pi}_N,\, \tilde{\pi}_N\}_P \approx 0, \quad \{\tilde{\pi}_N,\, \pi_i\}_P \approx 0,\quad
    \{\tilde{\pi}_N,\, \pi_\Lambda^{ij}\}_P \approx 0 \ , \nonumber  \\
    &\{\tilde{\pi}_N(\vx),\, \tilde{\pi}_h^{ij}(\vy)\}_P \approx \frac{1}{2}\fan^{y,ij}[\dxy] - \frac{1}{2}\cfan^{x,ij}[\dxy] = 0 \ .
\end{align}
Eq. (\ref{rel}) is used for the last equality in  Eq. (\ref{las}).
We can determine the Lagrange multipliers for $\pi_\Lambda^{ij}$ and $\tilde{\pi}_h^{ij}$ at this stage
since $\faa^{x,ij,kl}$ never vanishes as long as the Green's function $G_{\Lambda\Lambda,ij,kl}^{xy}$ exists.
Thus, their consistency conditions do not create any secondary constraints. On the other hand, the consistency conditions of $\tilde{\pi}_N$ and $\pi_i$ create 4 secondary constraints
\begin{align}
\label{}
     \dot{\tilde{\pi}}_N&= \{ \tilde{\pi}_N,\,H\}_P=\sqrt{h}(NF_N - F_\Lambda^{ij}\Lambda_{ij}) \equiv \chi\approx 0  \ , \\
    \dot{\pi}_i&=\{ \pi_i,\,H\}_P = -{\cal H}_i\approx 0 \ .
\end{align}
In general, the new constraint $\chi$ depends on the lapse function. The Poisson brackets of $\chi$ with $\tilde{\pi}_N$, $\pi_\Lambda^{ij}$ and $\tilde{\pi}_h^{ij}$ are obtained as
\begin{align}
\label{}
\{\chi(\vx),\,\tilde{\pi}_N(\vy)\}_P &= N\hat{F}^{x}_{NN}[\dxy]+\sqrt{h}F_N\dxy \nonumber \\
&\quad+\cfan^{y,ij}\left[ \int d^3x' G_{\Lambda\Lambda, ij,kl}^{yx'}\left( \sqrt{h}F_\Lambda^{kl}\delta^3(\vxx-\vx)-N\hat{F}_{N\Lambda}^{x, kl}[\delta^3(\vec{x}'-\vx)]
 \right)\right] \nonumber \\
&\equiv\hat{\chi}_N[\dxy]  \ , \nonumber \\
\label{}
\{\chi(\vx),\,\pi_\Lambda^{ij}(\vy)\}_P &= N\hat{F}^{x,ij}_{N\Lambda}[\dxy]-\sqrt{h}F_\Lambda^{ij}\dxy
            -\hat{F}_{\Lambda\Lambda}^{x, kl,ij}[\delta^3(\vx-\vy)]\Lambda_{kl} \nonumber \\
 &\equiv\hat{\chi}_\Lambda^{x,ij}[\dxy] \ , \nonumber \\
%
\label{}
   \{\chi(\vx),\,\tilde{\pi}_h^{ij}(\vy)\}_P &= N\hat{F}^{x,ij}_{Nh}[\dxy]
            -\hat{F}_{\Lambda h}^{x, kl,ij}[\delta^3(\vx-\vy)]\Lambda_{kl} \nonumber \\
 &\equiv\hat{\chi}_h^{x,ij}[\dxy] \ ,
\end{align}
where $\hat{\chi}_N$, $\hat{\chi}_\Lambda^{x,ij}$ and $\hat{\chi}_h^{x,ij}$ are derivative operators in general. $\hat{\chi}_N$ is not zero in general,
but should become weakly zero when the theory has the general covariance, see Appendix A.2 for this argument.
Introduce a linear combination of $\chi$ and $\tilde{\pi}^{ij}$ as
\begin{equation}
\label{ }
\chi'(\vx) \equiv \chi(\vx) + \hat{\chi}_\Lambda^{x,ij}\left[ \left< 2G_{\Lambda\Lambda,kl,ij}^{xx'}\,,\,\tilde{\pi}_h^{kl}(\vxx)\right>\right] \ ,
\end{equation}
we get
\begin{align}
\label{}
    \{\chi'(\vx),\,\pi_\Lambda^{ij}(\vy)\}_P &\approx \hat{\chi}_\Lambda^{x,ij}[\dxy] + \hat{\chi}_\Lambda^{x,kl}[-\mathbb{I}_{kl}^{ij}\dxy] = 0 \ , \nonumber \\
    \{\chi'(\vx),\,\tilde{\pi}_h^{ij}(\vy)\}_P &\approx \hat{\chi}_h^{x,ij}[\dxy] + 2\hat{\chi}_\Lambda^{x,mn}
    \left[ \int d^3x' G_{\Lambda\Lambda, kl,mn}^{xx'}\hat{\Pi}^{kl,ij}[\dxy]\right] \nonumber \\
    &\equiv \hat{\psi}^{x,ij}_h[\dxy] \ ,
\end{align}
where $\hat{\psi}^{x,ij}$ is a derivative operator.
We further introduce another linear combination of $\chi'$ and ${\pi_\Lambda^{ij}}$ as
\begin{equation}
\label{ }
\tc(\vx) \equiv \chi'(\vx) - \hat{\psi}^{x,ij}_h\left[\left<2G_{\Lambda\Lambda,ij,kl}^{x'x}\, ,\,\pi_\Lambda^{kl}(\vx)\right>\right] \ ,
\end{equation}
and we obtain
\begin{align}
\label{}
  &\{\tc(\vx),\,\tc(\vy)\}_P \approx 0,\quad \{\tc(\vx),\,\pi_i(\vy)\}_P \approx 0, \quad
    \{\tc(\vx),\,\pi_\Lambda^{ij}(\vy)\}_P \approx 0, \nonumber \\
    &\{\tc(\vx),\,\tilde{\pi}_h^{ij}(\vy)\}_P \approx \hat{\psi}_h^{x,ij}[\dxy] -\hat{\psi}^{x,kl}_h[\mathbb{I}^{ij}_{kl}\dxy] =0 \ , \nonumber \\
  &\{\tc(\vx),\,\tilde{\pi}_N(\vy)\}_P \approx \hat{\chi}_N[\dxy] \ .
\end{align}
%
%
Then, we consider the Poisson brackets with $\ch$.
We use the following relation for a functional $I$ which is invariant under the restricted spatial diffeomorphism $x^i\rightarrow x'^i(\vec{x})$,
\begin{equation}
\label{ }
\left\{\left<\ch, f^i\right>,\, I[h_{ij}, \pi^{ij}/\sqrt{h}, s, v^i, t_{ij}]\right\}_P \approx
      \int d^3x\left\{ \frac{\delta I}{\delta s}f^iD_is
      + \frac{\delta I}{\delta v^i}{\frak L}_fv^i +\frac{\delta I}{\delta t_{ij}}{\frak L}_ft_{ij}\right\} \ ,
\end{equation}
where $f^i$ is a general vector field. $s$ is a scalar, $v^i$ is a vector and $t_{ij}$ is a tensor under the restricted spatial diffeomorphism, and they do not depend on
both the conjugate momenta and $h_{ij}$.
This relation is an extended version of the one given in Appendix A of \cite{Mukohyama:2015gia} including the tensor field.
Using this relation, we find the nonzero Poisson brackets with $\ch$,
\begin{align}
\label{hh}
    \left \{\left<\ch, f^i\right>,\,\left<\tilde{\pi}_h^{kl}, T_{kl}\right>\right\}_P &\approx
   -\frac{1}{2}\int d^3x \cfan^{x,ij}[T_{ij}]f^kD_kN
   -\frac{1}{2}\int d^3x \cfaa^{x,ij,kl}[T_{ij}]{\frak L}_f\Lambda_{kl}  \ , \nonumber \\
    \left\{\left<\ch, f^i\right>,\,\left<\tilde{\chi}, f\right>\right\}_P &\approx
    \left\{\left<\ch, f^i\right>,\,\left<\chi', f\right>\right\}_P\approx
      \int d^3x\left\{ \frac{\delta\left< \chi', f\right>}{\delta N}f^iD_iN
      + \frac{\delta\left< \chi', f\right>}{\delta \Lambda_{ij}}{\frak L}_f\Lambda_{ij}\right\} \ ,
\end{align}
where $T_{ij}$ and $f$ are a general tensor field and a scalar field respectively. Taking notice that
\begin{equation}
\label{ }
\frac{\delta\left< \chi', f\right>}{\delta \Lambda_{ij}(\vx)} = \{ \left< \chi', f\right>,\,\pi_\Lambda^{ij}(\vx)\}
\approx 0\ ,
\end{equation}
the second equation in (\ref{hh}) eventually becomes
\begin{equation}
\label{ }
\left\{\left<\ch, f^i\right>,\,\left<\tilde{\chi}, f\right>\right\}_P \approx
        \int d^3x\left\{ \frac{\delta\left< \chi, f\right>}{\delta N}f^iD_iN
      \right\} \ .
\end{equation}
We define the following linear combination of the constraints
\begin{equation}
\label{ }
\mom \equiv \ch + \pi_ND_iN+\pi_\Lambda^{kl}D_i\Lambda_{kl}-2\sqrt{h}D_j\left(\frac{\pi_\Lambda^{jk}}{\sqrt{h}}\Lambda_{ik}\right) \ .
\end{equation}
Then, we can find all of the Poisson brackets of $\mom$ become zero
\begin{align}
\label{hhh}
   &\left \{\left<\mom, f^i\right>,\,\left<\tilde{\pi}_h^{kl}, T_{ij}\right>\right\}_P\approx
   \left \{\left<\ch, f^i\right>,\,\left<\tilde{\pi}_h^{kl}, T_{ij}\right>\right\}_P
    +\frac{1}{2}\int d^3x \cfan^{x,ij}[T_{ij}]f^kD_kN
   +\frac{1}{2}\int d^3x \cfaa^{x,ij,kl}[T_{ij}]{\frak L}_f\Lambda_{kl}
   \nonumber \\
   &\qquad\qquad\qquad\qquad\qquad\ \
   \approx0 \ , \nonumber \\
    &\left\{\left<\mom, f^i\right>,\,\left<\tilde{\chi}, f\right>\right\}_P\approx
    \left\{\left<\ch, f^i\right>,\,\left<\tilde{\chi}, f\right>\right\}_P
      -\int d^3x \left\{f^iD_iN\frac{\delta\left< \tc, f\right>}{\delta N}
      +{\frak L}_f\Lambda_{ij}\frac{\delta\left< \tc, f\right>}{\delta \Lambda_{ij}}\right\}
      \approx 0 \ , \nonumber \\
    &\left\{\left<\mom, f^i\right>,\,\left<\mom, g^i\right>\right\}_P\approx
    \left<\ch, {\frak L}_fg^i\right> \approx 0 \ ,\nonumber \\
   &\left\{\left<\mom, f^i\right>,\,\tilde{\pi}_N\right\}_P\approx 0,\quad
    \left\{\left<\mom, f^i\right>,\,\pi_j\right\}_P\approx 0,\quad
    \left\{\left<\mom, f^i\right>,\,\pi_\Lambda^{jk}\right\}_P\approx 0 \ ,
\end{align}
where $g^i$ is a general vector field, and we used the relations
\begin{equation}
\label{ }
      \frac{\delta\left< \tc, f\right>}{\delta N} \approx \frac{\delta\left< \chi, f\right>}{\delta N}, \quad
      \frac{\delta\left< \tc, f\right>}{\delta \Lambda_{ij}} \approx \frac{\delta\left< \chi', f\right>}{\delta \Lambda_{ij}} \approx 0 \ ,
\end{equation}
in the last weak equality in the second equation in (\ref{hhh}).
The consistency condition of $\tilde{\chi}$ just determines the Lagrange multiplier for $\tilde{\pi}_N$,
and that of $\mom$ generates no new constraint,
\begin{align}
\label{}
  \frac{d}{dt}\left< \mom, f^i\right>&\approx \left\{\left<\mom, f^i\right>,\,H\right\}_P \nonumber   \\
    &=\left<\ch, {\frak L}_fN^i\right>
    +\left\{\left<\ch, f^i\right>,\,\int d^3x'{\cal H}_{\bot}[h_{ij}, N, \Lambda_{ij}]\right\}_P
    \nonumber \\
    &\quad -\int d^3xd^3x'\left( f^iD_iN\frac{\delta {\cal H}_{\bot}(\vxx)}{\delta N(\vx)} +
      {\frak L}_f\Lambda_{ij}\frac{\delta {\cal H}_{\bot}(\vxx)}{\delta \Lambda_{ij}(\vx)} \right)  \nonumber \\
    &\approx 0  \ .
\end{align}
As a result, we find that the theory has the 6 first class constraints, $\pi_i$ and $\mom$, and the 14 second class constraints, $\tilde{\pi}_N$, $\tilde{\chi}$, $\pi_\Lambda^{ij}$ and $\tilde{\pi}_h^{ij}$ for the general functional $F$. Since the dimension of the phase space $(N,\,N^i,\,h_{ij}\,\Lambda_{ij},\,\pi_N,\,\pi_i,\,\pi_h^{ij},\,\pi_\Lambda^{ij})$ is 32, thus the number of degrees of freedom is
\begin{equation}
\label{ }
\frac{1}{2}( 32 - 2\times 6 - 14) = 3 \ .
\end{equation}
This is equal to the number of tensor modes in 3+1 dimensions plus a scalar mode.
If the theory has general covariance or other symmetries, the degrees of freedom will be less.
Therefore, we conclude that the XG theory has 2 tensor modes and 1 scalar mode at most,
and there do not appear any extra degrees of freedom making the theory unstable for all regions of the theory.

We have performed the Hamiltonian analysis of the XG theory without restoring the time diffeomorphism symmetry by the scalar field $\phi$. On the other hand, it has been recently pointed out that there appears an extra degree of freedom for some classes of the XG theory if we restore the time diffeomorphism symmetry and construct the general covariant form of the XG theory \cite{Langlois:2015cwa, Langlois:2015skt, Crisostomi:2016tcp,Crisostomi:2016czh}. Though we need to understand why such a difference occurs, our analysis ensures
that there does not appear any extra degree of freedom at least for the spatially covariant form of the XG theory (\ref{S}).

\section{Invariant transformations of the XG theory}\label{sec4}

In this section, we discuss the invariant transformation of the XG theory.
As we mentioned in Section \ref{sec1}, there are two
different ways to study the transformation which leaves the theory invariant.
One way is to search the transformation which preserves the forms of the action.
The other way is to search the transformation which preserves the forms of the Hamiltonian and the primary constraints.
The invariant transformation of the action belongs to a particular class of the point transformations,
and that of the Hamiltonian belongs to a particular class of the canonical transformations.
Moreover, the point transformation is a special case of the canonical transformation, and
it is easier to find the point transformation working in the Lagrangian formalism.
Depending on the problem, we will work with different formalisms.

We study the infinitesimal transformations first. If the infinitesimal transformations leave the theory invariant,
then the theory should be invariant also under the finite transformations which are obtained by repeating the infinitesimal transformations. Therefore, the infinitesimal invariant transformations can generate all the finite and smooth transformations which leave the theory invariant.
The metric transformations given in Section \ref{sec1} are smooth as long as they are invertible\footnote{There exist singular disformal transformations which can not be inverted. Through the singular transformations, the transformed theory
such as the mimetic theory \cite{Chamseddine:2013kea} may have
different structure of constraints and sometimes different number of degrees of freedom. }.
Therefore, the infinitesimal transformations we consider
in this article can include all the invertible classes of them. 
In this paper, we do not consider any discrete transformation
because it can not be generated from the infinitesimal transformations.

\subsection{Infinitesimal canonical transformation}

We shortly review the infinitesimal canonical transformation in this subsection \cite{Harashima}.
In general, a canonical transformation from a set of barred canonical
variables $(\bar{\varphi}_a, \bar{\pi}_a)$ to an unbarred set $(\varphi_a, \pi_a)$ is generated
by a generating functional ${\cal W}$. If ${\cal W}$ is given as the functional of $\bar{\varphi}_a$,
$\pi_a$ and the time $t$, the canonical transformation is given as
\begin{align}
\label{ct3}
    \bar{\pi}_a(\vx)&= \frac{\delta {\cal W}}{\delta \bar{\varphi}_a(\vx)} \ , \nonumber   \\
    \varphi_a(\vx)&= \frac{\delta {\cal W}}{\delta \pi_a(\vx)} \ , \nonumber \\
     H &= \bar{H} + \frac{\partial {\cal W}}{\partial t} \ ,
\end{align}
where $\bar{H}$ is the Hamiltonian of the barred system. On the other hand, the general point transformation is given as
\begin{equation}
\label{pt}
\varphi_a(\vx) = f_a[\bar{\varphi}(\vx), t] \ ,
\end{equation}
where $f_a$ are arbitrary but invertible functionals of $\bar{\varphi_a}$ and $t$.
Note that in general, we can not include the time derivatives of dynamical variables into the  arguments of $f_a$, otherwise we can not keep the equations of motion covariant.
If we consider ${\cal W} = \sum_{a}\int d^3x\,\pi_a f_a[\bar{\varphi}, t]$ as the generating functional of the canonical transformation, we obtain the point transformation in the Hamiltonian formalism
\begin{align}
\label{}
\varphi_a(\vx) &= f_a[\bar{\varphi}(\vx) , t] \ , \nonumber \\
\bar{\pi}_a(\vx) &= \int d^3x' \frac{\delta f_b}{\delta \bar{\varphi}_a(\vx)}\pi_b \ .
\end{align}
Therefore, the canonical transformation includes the point transformation.

Then, we consider the infinitesimal canonical transformation. First, we consider
the identity transformation which is generated by ${\cal I} = \sum_a\int d^3x\, \pi_a \bar{\varphi}_a$,
\begin{align}
\label{}
\bar{\pi}_a&= \frac{\delta {\cal I}}{\delta \bar{\varphi}_a} = \pi_a \ , \nonumber   \\
    \varphi_a&= \frac{\delta {\cal I}}{\delta \pi_a} = \bar{\varphi}_a \ .
\end{align}
The infinitesimal canonical transformation is obtained by taking the infinitesimal variation from the identity transformation. Defining $\epsilon$ as the infinitesimal parameter, we give the generating functional of infinitesimal transformation as
\begin{equation}
\label{ }
{\cal W} = \sum_a\int d^3x\, \left( \pi_a \bar{\varphi}_a + \epsilon\,{\cal G}[\pi, \bar{\varphi}, t]\right) \ ,
\end{equation}
and we obtain the infinitesimal canonical transformation
\begin{align}
\label{}
    \bar{\pi}_a& = \pi_a + \epsilon\frac{\delta \int {\cal G}d^3x }{\delta \bar{\varphi}_a} \ , \nonumber   \\
    \varphi_a&= \bar{\varphi}_a + \epsilon \frac{\delta \int  {\cal G}d^3x}{\delta \pi_a} \ .
\end{align}
We will call the functional ${\cal G}$ the generator of infinitesimal canonical transformation.
We need only up to the first order of $\epsilon$ for all of the quantities so that we can replace the argument $\bar{\varphi}_a$ in ${\cal G}$ to $\varphi_a$. Consequently, we obtain the following infinitesimal canonical transformation generated by ${\cal G}[\pi, \varphi, t]$:
\begin{align}
\label{ict}
    \bar{\pi}_a&= \pi_a + \epsilon \frac{\delta \int {\cal G}d^3x'}{\delta \varphi_a}\equiv \pi_a +\epsilon ({\cal G})_{\varphi_a}  \ , \nonumber \\
    \bar{\varphi}_a&= \varphi_a - \epsilon\frac{\delta \int {\cal G}d^3x' }{\delta \pi_a} \equiv \varphi_a -\epsilon({\cal G})_{\pi_a} \ , \nonumber \\
   H &= \bar{H} + \epsilon \frac{\partial \int {\cal G}d^3x'}{\partial t} \equiv \bar{H} +\epsilon ({\cal G})_t \ .
\end{align}
The bracket with the index denotes the derivatives of the quantities in the bracket with
respect to the index, $(\, )_N\equiv \int d^3x' \delta /\delta N$.
We note that we can always invert the infinitesimal transformation up to the first order of $\epsilon$.

\subsection{Invariant infinitesimal point transformation}\label{sub4}

In this subsection, we derive the infinitesimal point transformation which leaves the XG theory invariant.
We work with the Lagrangian formalism.  We start from the original XG action (\ref{S}) with the barred variables
\begin{equation}
\label{bxg}
\bar{S} = \int dtd^3x \bN\sqrt{\bh}\bar{F}[\bN, \,\bh_{ij},\, \bK_{ij},\, \bar{R}_{ij},\,\bar{D}_i,\, t] \ ,
\end{equation}
where $\bar{D}_i$ is the spatially covariant derivative with respect to $\bh_{ij}$.
The infinitesimal point transformation is obtained by taking the infinitesimal form of Eq. (\ref{pt}),
\begin{equation}
\label{ipt}
\bN = N-\epsilon P_N[N,N^i,h_{ij},t],\quad \bN^i = N^i-\epsilon P^i[N,N^i,h_{ij},t],
\quad \bh_{ij} = h_{ij} -\epsilon P_{ij}[N,N^i,h_{ij},t] \ .
\end{equation}
where $P_a\equiv(P_N,\,P^i,\,P_{ij})$ are arbitrary functionals. We omit to write $R_{ij}$, $D_i$ etc. in the arguments of $P_a$.
The infinitesimal transformation of the extrinsic curvature is obtained from Eq. (\ref{ipt}) as
\begin{align}
\label{Pk}
\bK_{ij} &= K_{ij} -\epsilon {P_{K}}_{ij} \ , \nonumber \\
{P_{K}}_{ij} &= -\frac{P_N}{N}K_{ij} +\frac{1}{2N}\left\{ \dot{P}_{ij} -2D_{(i}(P_{j)k}N^k + P_{j)})
+ N^k(2D_{(i}P_{j)k}-D_kP_{ij})\right\} \ ,
\end{align}
where $D_{(i}P_{j)}=(D_iP_j+D_jP_i)/2$.
After the infinitesimal transformation, the action up to the first order of $\epsilon$ becomes
\begin{align}
\label{S'}
\bar{S} &= S'= \int dtd^3x N\sqrt{h}\left[ F - \epsilon\left\{ F_NP_N + F_h^{ij}P_{ij} + F_{K}^{ij}P_{Kij}\right\}\right] \ , \\
\label{F}
F&= \bar{F}[N,\, h_{ij}, \, K_{ij},\, R_{ij},\,D_i,\,t] \ ,
\end{align}
where $F$ has the same form of functional as $\bar{F}$ with the unbarred arguments, and $F_{K}^{ij}$ is obtained from $F_\Lambda^{ij}$ by replacing $\Lambda_{ij}$ with $K_{ij}$. Note that we treat $K_{ij}$ and $h_{ij}$ as independent variables. 

We aim to leave the form of the action invariant,
\begin{equation}
\label{Sp}
S' = \int dt d^3x N\sqrt{h}F'[N,\, h_{ij}, \, K_{ij},\, R_{ij},\,D_i,\,t] \ .
\end{equation}
To achieve this, we must constrain the functionals $P_a$ in the action (\ref{S'}). First, we consider how to constrain the functional $P_{ij}$.
Since it has $\dot P_{ij}$ in $P_{Kij}$, $P_{ij}$ should
not depend on $N$ and $N^i$ so that $\dot{N}$ and $\dot{N}^i$ do not appear in the action.
We may be able to eliminate those time derivatives as total derivatives in $F_{K}^{ij}$. It is difficult, however,
to change them into total derivatives for the general functional $F$. Thus, we require $P_{ij}$ to depend only on $h_{ij}$ and $t$.
A possible and simple candidate for $P_{ij}$\footnote{We may use other functions
 such as $R_{ij}$ for $P_{ij}$. However, the simplest form is given as (\ref{Pij}) and we consider only the simplest form in this article.} is
\begin{equation}
\label{Pij}
P_{ij} = a(t)h_{ij}\ ,
\end{equation}
where $a(t)$ is an arbitrary function of $t$. Combining Eqs. \eqref{Pk} and (\ref{Pij}), we get
\begin{equation}
\label{pk2}
P_{Kij} = \left(a-\frac{P_N}{N}\right)K_{ij} + \frac{\dot{a}}{2N}h_{ij} -\frac{1}{N}D_{(i}P_{j)} \ .
\end{equation}
If $P_N$ and $P^i$ do not depend on $N^i$, then the action (\ref{S'}) has the desired form (\ref{Sp})
since all the terms are constructed with the building blocks only. So the condition is
\begin{align}
\label{sh}
0&=\frac{\delta}{\delta N^k}\int dtd^3x \sqrt{h}\left\{\left(NF_N - F_K^{ij}K_{ij}\right)P_N -F_K^{ij}D_iP_j\right\} \nonumber \\
 &= \int dtd^3x \sqrt{h}\left\{ \left(NF_N - F_K^{ij}K_{ij}\right)\frac{\delta P_N}{\delta N^k} + D_iF_K^{ij}\frac{\delta P_j}{\delta N^k}\right\} \ .
\end{align}
If the functional $F$ has up to the $n$-th order of $K_{ij}$, then the coefficient of $P_N$ in Eq. (\ref{sh}) is in the $n$-th order of $K_{ij}$,
and that of $P^i$ is in the $(n-1)$-th order of $K_{ij}$. Since $P_N$ and $P^i$ do not depend on $K_{ij}$, the condition (\ref{sh}) tells us that
\begin{equation}
\label{ }
\frac{\delta P_N}{\delta N^i} = 0 ,\quad \frac{\delta P_j}{\delta N^i} = 0\ .
\end{equation}
As a result, the XG theory is invariant under the infinitesimal point transformation
\begin{equation}
\label{ipt2}
\bN = N -\epsilon P_N[N, h_{ij}, t],\quad \bN^i = N^i-\epsilon P^i[N, h_{ij}, t],\quad \bh_{ij}= h_{ij}-\epsilon a(t)h_{ij} \ .
\end{equation}
The action transforms as
\begin{align}
\label{SF}
\bar{S} &= S' \nonumber \\
           &=\int dtd^3x \bN\sqrt{\bh}\bar{F}[\bN,\bh_{ij},\bK_{ij}, \bar{R}_{ij}, \bar{D}_i,t]
                 \nonumber \\
           &=  \int dtd^3x N\sqrt{h}F'[N, h_{ij}, K_{ij}, R_{ij}, D_i, t] \ , \nonumber  \\
       F' &=  F - \epsilon\left\{ F_NP_N + aF_h + \left(a- \frac{P_N}{N}\right)F_{K}^{ij}K_{ij}
         +\frac{\dot{a}}{2N}F_K-\frac{1}{N}F_K^{ij} D_iP_j \right\} \ ,
\end{align}
where $F_h\equiv h_{ij}F_h^{ij}$.
The form of action is preserved by replacing $\bar{F}$ with $F'$.

We can immediately promote the infinitesimal transformation (\ref{ipt2}) to finite transformation by removing the infinitesimal parameter $\epsilon$.
The finite transformation is
\begin{align}
\label{lt}
    &\bN= N - P_N[N, h_{ij}, t], \quad
    \bN^i = N^i - P^i[N, h_{ij}, t],\quad
    \bh_{ij}= {\cal A}(t)h_{ij} \ , \nonumber \\
    &\bar{\Gamma}^i_{jk}= \Gamma^i_{jk}, \quad \bar{R}_{ij} = R_{ij}\ , \nonumber \\
    &\bar{K}_{ij}= \frac{{\cal A}N}{\bN}\left( K_{ij} + \frac{\dot{{\cal A}}}{2{\cal A}N}h_{ij} - D_{(i}P_{j)}\right)\ .
\end{align}
No extra variable other than the building blocks is introduced by the finite transformation, so the action is invariant under the finite transformation (\ref{lt}) as long as the transformation is invertible.
Restoring the full diffeomorphism by using the Stuck\"{e}lberg field $\phi$,
we may derive the metric transformation of the theory. Under the transformation (\ref{lt}),
the components of the spacetime metric $\bar{g}_{\mu\nu}$ transform as
\begin{align}
\label{gb}
    \bar{g}_{00}&= -\bN^2 + \bN^i\bN_i = -(N-P_N)^2 + {\cal A}(N^i-P^i)(N_i-P_i) \ ,  \nonumber \\
    \bar{g}_{0i}&= \bN_{i} = {\cal A}(N_i-P_i) \ , \nonumber \\
    \bar{g}_{ij}&= \bh_{ij} = {\cal A}h_{ij}  \ .
\end{align}
From these results, we expect that the transformation for the spacetime metric include the general covariant form of $P^i$ up to the quadratic order. We define the general covariant form of $P^i$ with a 4-vector $V^\mu$ as
\begin{align}
\label{}
    P^\mu&\equiv h^\mu_\nu V^\nu \ , \nonumber  \\
    P^0 &= 0,\quad P^i = V^i,\quad P_0= N_iP^i,\quad P_i= V_i\ .
\end{align}
So the transformation of $g_{\mu\nu}$ is
\begin{equation}
\label{disform70}
\bar{g}_{\mu\nu} = {\cal A}[\phi]g_{\mu\nu} + b\nabla_\mu\phi\nabla_\nu\phi + c\nabla_{(\mu}\phi P_{\nu)} + d_1(P^\rho P_\rho)\nabla_\mu\phi\nabla_\nu\phi + d_2P_\mu P_\nu \ ,
\end{equation}
where the coefficients $b,\,c,\,d_1$ and $d_2$ are functionals of $\phi,\,g_{\mu\nu}$ and $\nabla_{\mu}$.
We neglect the term $P^\rho \nabla_{\rho}\phi P_{(\mu}\nabla_{\nu)}\phi$ since it vanishes in the unitary gauge $\phi= t$.
Comparing Eq. \eqref{disform70} with Eq. (\ref{gb}), in the unitary gauge we get
\begin{equation}
\label{ }
b= \frac{{\cal A}}{X} - \left( X^{-1/2} -P_N\right)^2,\quad c= -2{\cal A},\quad d_1={\cal A},\quad d_2=0 \ .
\end{equation}
Since
\begin{equation}
\label{ }
P_\mu = V_\mu + n_\mu n^\nu V_\nu = V_\mu + V^\nu\nabla_\nu\phi \frac{\nabla_\mu \phi}{X} \ ,
\end{equation}
so
\begin{align}
\label{glt}
\bar{g}_{\mu\nu} &= {\cal A}[\phi]g_{\mu\nu} + {\cal B}\nabla_\mu\phi\nabla_\nu\phi -2{\cal A}[\phi]\nabla_{(\mu}\phi V_{\nu)} \ , \nonumber \\
{\cal B}& = b + {\cal A}\left\{ V^\mu V_\mu+ \frac{(V^\rho\nabla_\rho\phi)^2}{X} - \frac{2V^\rho\nabla_\rho\phi}{X}\right\} \ ,.
\end{align}
The metric transformation \eqref{glt} leaves the XG theory invariant in the unitary gauge,
and it is more general than the disformal transformation (\ref{disf}). Note that the form of $V^\mu$ and $b=b[\phi,\, g_{\mu\nu},\,\nabla_\mu]$ are constrained as they have the form of $P_N$ and $P^i$ in (\ref{lt}) in the unitary gauge.
If we choose $V^\mu= 0$ and $b= b[\phi, X]$, then the transformation (\ref{glt}) reduces to
the disformal transformation (\ref{disf}). To see how the transformation (\ref{glt}) is more general,
as a simple example, we consider the transformation generated by $P_i = D_i(1/N^2)$. The 4-vector $V^\mu$ is
\begin{align}
\label{ }
P_\mu &= D_\mu\left(\frac{1}{N^2}\right) = h^\nu_\mu\nabla_\nu X  = h^\nu_\mu V_\nu \ , \nonumber \\
V_\mu &= \nabla_\mu X \ .
\end{align}
Thus, the transformation \eqref{glt} which leaves the XG theory invariant contains higher order derivatives through the vector product $V^{\mu}$ and $b$.

The general covariant form of the XG theory should be invariant under the invariant metric transformation (\ref{glt}). As we mentioned in the previous section, however, the stability of the general covariant form of the XG theory is not ensured. Thus, we must be careful of treating the invariant metric transformation in the general covariant form. 
Although it is uncertain whether the general covariant form of the XG theory is stable or not, but the spatially covariant form of the XG theory (\ref{bxg}) is invariant under the invariant point transformation (\ref{lt}), so the stability of the theory is ensured since the spatially covariant form of the XG theory has at most 3 degrees of freedom.

\subsection{Invariant infinitesimal canonical transformation}\label{sub43}

In this subsection, we derive the infinitesimal canonical transformation which leaves the XG theory invariant,
and we work in the Hamiltonian formalism.
The generator of the infinitesimal point transformation in the last subsection is
\begin{equation}
\label{gp}
{\cal G}_P = \int d^3x\left(\pi_N P_N + \pi_i P^i + a\pi_h + \pi_\Lambda^{ij}P_{\Lambda ij}\right) \ ,
\end{equation}
where $P_{\Lambda ij}$ has the same form as $P_{Kij}$ in Eq. (\ref{pk2}) with the argument $\Lambda_{ij}$ instead of $K_{ij}$,
\begin{equation}
\label{paij}
P_{\Lambda ij} = \left(a-\frac{P_N}{N}\right)\Lambda_{ij} + \frac{\dot{a}}{2N}h_{ij} -\frac{1}{N}D_{(i}P_{j)} \ .
\end{equation}

The generator ${\cal G}_P$ of the point transformation is linear in the momenta. The generators of canonical transformations, however,
usually include nonlinear momenta terms.  In this paper,  we consider only the infinitesimal canonical transformation
which contains up to second order in the momenta for simplicity.

We obtained the primary constraints (\ref{pri}) of the XG theory in the previous section. As we mentioned at the beginning of this section,
we must leave not only the form of the Hamiltonian but also the structure of primary constraints invariant under the canonical transformation. Namely, we need
\begin{equation}
\label{ }
(\bp_N,\,\bp_i,\,\bp_\Lambda^{ij},\,\bar{\tilde{\pi}}_h^{ij}) \approx 0\quad \rightarrow \quad (\pi_N,\,\pi_i,\,\pi_\Lambda^{ij},\,\tilde{\pi}_h^{ij}) \approx 0 \ .
\end{equation}
%
The momenta except $\pi_h^{ij}$ should vanish weakly after the transformation.
If we take any functional which contains quadratic momenta except $\pi_h^{ij}$ terms as the generator,
all of the variations generated by them will vanish weakly since the first order terms of $\epsilon$ in Eq. (\ref{ict}) vanish weakly for that generator.
Therefore, it is trivial to consider the generator which contains the quadratic momenta except $\pi_h^{ij}$ terms.
On the other hand, the momentum $\pi_h^{ij}$ is not the primary constraint, and thus we can construct a nontrivial generator by using it.
We consider a functional which contains more than one $\pi_h^{ij}$ and another functional which does not contain any momenta in the generator.
Hence, we express the generator of the infinitesimal canonical transformation which includes the infinitesimal point transformation as
\begin{equation}
\label{gc}
{\cal G}= {\cal G}_P + \int d^3x \left\{\sum_a C_{\pi_a}[\pi_a, \pi_h^{ij}; N, N^i, h_{ij}, \Lambda_{ij}, t] + C[N,N^i, h_{ij},\Lambda_{ij}, t]\right\} \ ,
\end{equation}
where $C_{\pi_a}=\hat{B}_{aij}[N,N^i, h_{ij},\Lambda_{ij}, t]\pi_a\pi^{ij}_h$ are the scalar quantities bi-linear in $(\pi_a, \pi_h^{ij})$, and we again omit to write $R_{ij}$, $D_i$ etc. in the arguments of $\hat{B}_{a ij}$ and $C$. In general, $\hat{B}_{aij}$ is a derivative operator.
Now we study the forms of the functionals $C_{\pi_a}$ and $C$ which leave the XG theory invariant.

Under the canonical transformation, $H_{tot}$ becomes,
\begin{align}
\label{ }
H_{tot}&= \bar{H}_{tot} + \epsilon ({\cal G})_t \nonumber \\
              &= \bar{H} + \int d^3x\left( \bar{\lambda}_N\bp_N +
               \bar{\lambda}^i\bp_i + \bar{\lambda}_{\Lambda ij}\bp_\Lambda^{ij} + \bar{\lambda}_{hij}\tbp_h^{ij} \right) + \epsilon ({\cal G})_t \nonumber \\
    &= H
    + \int d^3x\left( \lambda_N\pi_N +
               \lambda^i\pi_i + {\lambda}_{\Lambda ij}\pi_\Lambda^{ij} + {\lambda}_{hij}\tilde{\pi}_h^{ij} \right) \ .\end{align}
In order to keep the form of $H_{tot}$, the primary constraints $\bar{\pi}_a$ must transform to the linear combination of the primary constraints $\pi_a$.
Using the generator (\ref{gc}), we get
\begin{align}
\label{pa}
\bp_\Lambda^{ij} &= \pi_\Lambda^{ij} + \epsilon \left\{ (P_{\Lambda kl})_\Lambda^{ij}[\pi_\Lambda^{kl}]+(\cn)_\Lambda^{ij} + (\ci)_\Lambda^{ij} + (\ca)_\Lambda^{ij} + (\cph)_\Lambda^{ij} + (C)_\Lambda^{ij}               \right\} \nonumber \\
          &\approx \pi_\Lambda^{ij} + \epsilon\left\{ (\cph)_\Lambda^{ij} + (C)_\Lambda^{ij}\right\} \nonumber \\
          &\approx 0 \ .
\end{align}
The transformed constraint includes not only $\pi_\Lambda^{ij}$ but also the other primary constraints through $C_{\pi_a}$ except $C_{\pi_h}$
and the term
which does not contain any constraint. 
Although the other primary constraints appear after the transformation,
we can always preserve the structures of the primary constraints in $H_{tot}$ by redefinition of the multipliers.
Therefore, the primary constraints on the right hand side of the first equality in Eq. (\ref{pa}) vanish weakly.
However, we can not eliminate the terms which do not contain any constraint. 
The functionals $C_{\pi_h}$ and $C$ must not create such terms in Eq. \eqref{pa},
and we will discuss the constraints later. 

Under the infinitesimal canonical transforation, the constraints $\bp_N$ and $\bp_i$ transform to
\begin{align}
\label{pn}
    \bp_N&\approx \pi_N + \epsilon \left\{ (C_{\pi_h})_N + (C)_N\right\} \approx 0 \ , \\
    \label{pi}
    \bp_i &\approx \pi_i +\epsilon \left\{ (\cph)_{Ni} + (C)_{Ni}\right\} \approx 0  \ .
    \end{align}
The remaining constraint $\tbp_h^{ij}$ transforms as
\begin{equation}
\label{ }
\tbp_h^{ij} = \bar{\pi}_h^{ij} - \frac{\sqrt{\bar{h}}}{2}\bar{F}^{ij}_{\bar{\Lambda}} =\bar{\pi}_h^{ij} - \frac{1}{2\bN}\int d^3x'\frac{\delta(\bN\sqrt{\bar{h}}\bar{F})(\vxx)}{\delta \bar{\Lambda}_{ij}}  \approx 0 \ .
\end{equation}
Introducing the functional $F'$,
\begin{equation}
\label{F'}
F' \equiv \frac{\bN\sqrt{\bar{h}}}{N\sqrt{h}}\bar{F} = F -\epsilon\left\{F_N({\cal G})_{\pi_N}
+F_h^{ij}({\cal G})_{\pi_hij} + F_\Lambda^{ij}({\cal G})_{\pi_\Lambda ij}\right\} \ ,
\end{equation}
where $F$ is given in (\ref{F}) replacing $K_{ij}$ to $\Lambda_{ij}$, we get
\begin{align}
\label{fr}
    \int d^3x'\frac{\delta(\bar{N}\sqrt{\bar{h}}\bar{F})(\vxx)}{\delta \bar{\Lambda}_{ij}} &= \sum_a\int d^3x'd^3y\left(\frac{\delta \varphi_a(\vy)}{\delta \bar{\Lambda}_{ij}}\frac{\delta (N\sqrt{h}F')(\vxx)}{\delta \varphi_a(\vy)} +\frac{\delta \pi_a(\vy)}{\delta \bar{\Lambda}_{ij}}\frac{\delta (N\sqrt{h}F')(\vxx)}{\delta \pi_a(\vy)}\right) \nonumber \\
      &= N\sqrt{h}F_\Lambda^{'ij} +\epsilon\,\sum_a\int d^3x'd^3y \frac{\delta ({\cal G})_{\pi_a}(\vy)}{\delta \Lambda_{ij}}\frac{\delta (N\sqrt{h}F')(\vxx)}{\delta \varphi_a(\vy)} + {\cal O}(\epsilon^2) \ .
\end{align}
For simplicity, we assume
\begin{equation}
\label{as}
\frac{\delta ({\cal G})_{\pi_N}}{\delta \Lambda_{ij}} \approx 0,\quad \frac{\delta ({\cal G})_{\pi k}}{\delta \Lambda_{ij}} \approx 0,\quad \frac{\delta ({\cal G})_{\pi_h kl}}{\delta \Lambda_{ij}} \approx 0 \ .
\end{equation}
Note that these assumptions are automatically satisfied for the generator \eqref{gp}, so they are satisfied for the infinitesimal point transformation derived in the previous subsection.
Under these assumptions, we obtain
\begin{align}
\label{ph}
\tbp_h^{ij} &\approx  {\pi}_h^{ij}-\frac{\sqrt{h}}{2}F_\Lambda^{'ij} + \epsilon\left\{ a\pi_h^{ij} +(\cph)_h^{ij} + (C)_h^{ij} -\frac{P_N+ (\cn )_{\pi_N}}{2N}\sqrt{h}F_\Lambda^{'ij} \right. \nonumber \\
                    &\qquad \qquad\left.-\frac{1}{2N}\int  d^3x'd^3y\frac{\delta \left(P_{\Lambda kl}+ (\ca)_{\pi_\Lambda kl}\right)(\vy)}
                      {\delta \Lambda_{ij}} \frac{\delta (N\sqrt{h}F')(\vxx)}{\delta \Lambda_{kl}(\vy)}\right\} \nonumber \\
                      &\approx 0 \ .
\end{align}
Using Eqs. (\ref{F'}) and (\ref{as}), we obtain
\begin{align}
\label{th}
    H= \bar{H} + \epsilon\frac{\partial {\cal G}}{\partial t}
    &= \int d^3x \left[ N^i\ch + N\sqrt{h}\left(F_\Lambda^{'ij}\Lambda_{ij}-F'\right)+ \epsilon\,\delta {\cal H}\right] \nonumber \\
    &+ (\text{linear combinations of the primary constraints}) \ , \\
    \label{nh}
    \delta {\cal H} &= \delta (N^i\ch) + \delta {\cal H}_{\bot} \ , \\
   \delta (N^i\ch) &\equiv 2\left\{ \left( {(\cph)_h}^i_j + {(C)_h}^i_j- \pi_h^{ik}(\cph)_{\pi_hjk}\right) D_iN^j \right. \nonumber \\
   &\quad\qquad\left.-{\pi_h}^i_jD_i\left(P^j+(C_\pi)_\pi^j\right) -\frac{1}{2}\pi_h^{ik}N^jD_j(\cph)_{\pi_hik} \right\}\ ,  \\
   \label{dhp}
   \delta {\cal H}_\bot &\equiv  (C_{\pi_h})_t+(C)_t + \dot{a}\pi_h -N\sqrt{h}F_\Lambda^{'ij}\left( P_{\Lambda ij} + (C_{\pi_\Lambda})_{\pi_\Lambda ij}\right) \nonumber \\
   &\quad\qquad +\Lambda_{ij}\int  d^3x'd^3y\frac{\delta \left(P_{\Lambda kl}+ (\ca)_{\pi_\Lambda kl}\right)(\vy)}
                      {\delta \Lambda_{ij}} \frac{\delta (N\sqrt{h}F')(\vxx)}{\delta \Lambda_{kl}(\vy)}\ .
\end{align}
The linear combinations of the primary constraints can be absorbed into $H_{tot}$ by the redefinition of the multipliers.

Before we study the constraints on $C_{\pi_a}$ and $C$, we first discuss the point transformations and
confirm that the functional $P_{\Lambda ij}$ is given by Eq. (\ref{paij})
in the Hamiltonian formalism. Setting $C_{\pi_a}$ and $C$ to zero, and
using Eq. (\ref{paij}), we get
%
\begin{align}
\label{}
    &\bp_N \approx \pi_N\approx 0,\quad \bp_i \approx \pi_i\approx 0,\quad \bp_\Lambda^{ij} \approx \pi_\Lambda^{ij}\approx 0 \ ,\nonumber \\
    &\tbp_h^{ij} \approx \pi_h^{ij} -\frac{\sqrt{h}}{2}F_\Lambda^{'ij}
           + \epsilon\, a\left( \pi_h^{ij} - \frac{\sqrt{h}}{2}F_\Lambda^{'ij}\right)\approx 0\ ,\nonumber \\
&H \approx \int d^3x \left[ N^i\ch + N\sqrt{h}\left( F_\Lambda^{'ij}\Lambda_{ij} - F'\right)
         +\epsilon\,\dot{a}h_{ij}\left(\pi_h^{ij} -\frac{\sqrt{h}}{2}F_\Lambda^{'ij}\right)\right] \ .
\end{align}
The first order terms of $\epsilon$ are proportional to the primary constraint $\pi_h^{ij}-\frac{\sqrt{h}}{2}F_\Lambda^{'ij}$ in the unbarred system by replacing $\bar{F}$ with $F'$,
and they are absorbed into $H_{tot}$ by the redefinition of the multipliers.
The functional $F'$ defined by Eq. (\ref{F'}) now equals to $F'$ in the action (\ref{SF}) by replacing $K_{ij}$ with $\Lambda_{ij}$.
Therefore, the result in the Hamiltonian formalism completely agrees with that in the Lagrangian formalism obtained in the previous subsection,
and we obtain $P_{\Lambda ij}$ from  $P_{K ij}$  by replacing $K_{ij}$ with $\Lambda_{ij}$.
From the above analysis, we can find that the invariant infinitesimal point transformation in the Hamiltonian formalism
is closed only by itself. 

Now we discuss the functional forms of $C_{\pi_h}$ and $C$. From the assumptions (\ref{as}), we get $(C_{\pi_h})_\Lambda^{ij}\approx 0$.
Combining this result with Eq. (\ref{pa}), it requires that $(C)_\Lambda^{ij}\approx 0$. Since any combination of $C_{\pi_h}$ and $C$
can not be proportional to the primary constraint $\tilde{\pi}_h^{ij}$, so they are independent of $\Lambda_{ij}$.
From the conditions (\ref{pn}) and (\ref{pi}), we find that $C_{\pi_h}$ and $C$ are also independent of $N$ and $N_i$, so
\begin{align}
\label{}
&C_{\pi_h}= C_{\pi_h}[\pi_h^{ij}, \pi_h^{kl}; h_{ij}, t] \ , \nonumber \\
    &C= C[h_{ij}, t]  \ .
\end{align}
%

To prohibit $\delta \cal H$ from depending on $N^i$,
we must take the functional $C$ as
\begin{equation}
\label{Cm}
C=\sqrt{h}(c_1(t)+ c_2(t)R) \ ,
\end{equation}
where $c_1$ and $c_2$ are arbitrary functions of time. See Appendix B for the derivation of (\ref{Cm}).

For simplicity, we choose $\cph$ by hand as
\begin{equation}
\label{cph}
\cph = c_3(t) \frac{\pi_h^2}{\sqrt{h}} + c_4(t)\frac{\pi_h^{ij}\pi_{hij}}{\sqrt{h}} \ ,
\end{equation}
and we do not consider other more complicated functionals.
Once we determine the functional $\cph$, we obtain $C_{\pi_\Lambda}$ as
\begin{align}
\label{ca}
    &\ca = c_3(t)\frac{\pi_h}{\sqrt{h}}\left( 2\pi_\Lambda^{ij}\Lambda_{ij} - \frac{1}{2}\pi_\Lambda\Lambda \right) + c_4(t)\frac{\pi_{hij}}{\sqrt{h}} \left( 2\pi_\Lambda^{jk}\Lambda_{k}^i - \frac{1}{2}\pi_\Lambda^{ij}\Lambda \right)
     + c_5 \left( \frac{\pi_h\pi_\Lambda^{ij}}{\sqrt{h}}\Lambda_{ij} - \frac{\pi_h^{ij}\pi_\Lambda}{\sqrt{h}}\Lambda_{ij}\right) \nonumber \\
     &\qquad+\frac{1}{2N}\left( \dot{c}_3\frac{\pi_h\pi_\Lambda}{\sqrt{h}} + \dot{c}_4\frac{\pi_{hij}\pi_\Lambda^{ij}}{\sqrt{h}}\right) -\frac{(\cn)_{\pi_N}}{N}\pi_\Lambda^{ij}\Lambda_{ij} - \frac{1}{N}\pi_\Lambda^{ij}D_i(\ci)_{\pi j}\ ,
\end{align}
where the arbitrary functional $c_5\equiv c_5[N, N^i, h_{ij}, t]$.
By introducing the functional $F''$, we rewrite the Hamiltonian (\ref{th}) as
\begin{align}
\label{H'}
    H &\approx \int d^3x \left[ N^i{\cal H}_i  + N\sqrt{h}\left( F_\Lambda^{''ij}\Lambda_{ij} - F''\right)\right] \ ,  \\
    \label{tpih}
    \tilde{\bar{\pi}}_h^{ij}&\approx  \tilde{\pi}_h^{ij} = \pi_h^{ij} - \frac{\sqrt{h}}{2}F_\Lambda^{''ij} \approx 0 \ ,  \\
    \label{F''}
    F'' &= F -\epsilon\left\{ F_N\left( P_N+(\cn)_{\pi_N}\right) + F_h^{ij}\left( ah_{ij} +(\cph)_{\pi_hij}\right) + F_\Lambda^{ij}\left( P_{\Lambda^{ij}}+(\ca)_{\pi_\Lambda ij}\right) \right. \nonumber \\
       &\left. \quad+ c_1\Lambda - 2c_2G^{ij}\Lambda_{ij} +\frac{\dot{c_1}}{N} +\frac{\dot{c_2}}{N}R\right\} \ .
\end{align}
Again, see Appendix B for the detailed derivations of (\ref{ca})--(\ref{F''}).
Although the momentum $\pi_h^{ij}$ appears in $F''$ through $C_{\pi_a}$, we can rewrite it to $\sqrt{h}F_\Lambda^{''ij}/2$ using the weak equality (\ref{tpih}),
and therefore the form of Hamiltonian is preserved after the transformation.

%
%

At last, we consider whether $\cn$ and $\ci$ can depend on $N^i$ or not.
To leave the Hamiltonian invariant, the functional $F''$ in the transformed Hamiltonian $H$ should not depend on $N^i$.
We extract the terms which possibly depend on $N^i$ from $F''$
\begin{equation}
\label{ }
\int d^3x N\sqrt{h}F'' 
    \supset \epsilon\int d^3x \sqrt{h}\left\{ \left( NF_N-F_\Lambda^{ij}\Lambda_{ij}\right)(\cn)_{\pi_N} -F_\Lambda^{ij}D_i(\ci)_{\pi j} \right\}  \ .
\end{equation}
To prohibit $F''$ from depending on $N^i$, we require that
%
\begin{equation}
\label{cnn}
\frac{\delta}{\delta N^i} \int d^3x \sqrt{h}\left\{ \left( NF_N-F_\Lambda^{ij}\Lambda_{ij}\right)(\cn)_{\pi_N} -F_\Lambda^{ij}D_i(\ci)_{\pi j} \right\}  \approx 0 \ .
\end{equation}
As in the case of infinitesimal point transformation, we find that $\cn$ and $\ci$ are independent of $N^i$.
If the functional $F$ has up to the $n$-th power of $\Lambda_{ij}$, the $\cn$ term in Eq. (\ref{cnn}) is $(2n-1)$-th power of $\Lambda_{ij}$
since $(C_{\pi_a})_{\pi_a}$ has $(n-1)$-th power of $\Lambda_{ij}$ using the weak equality (\ref{tpih}). On the other hand,
the $\ci$ term is $(2n-2)$-th power of $\Lambda_{ij}$. Therefore, in order to satisfy the condition (\ref{cnn}) for general functional $F$, we need
\begin{equation}
\label{ }
\frac{\delta (\cn)_{\pi_N}}{\delta N^i} = 0,\quad \frac{\delta (\ci)_{\pi j}}{\delta N^i} = 0\ .
\end{equation}

In summary, in addition to the infinitesimal point transformation generated by ${\cal G}_P$ (\ref{gp}),
we obtain the infinitesimal invariant canonical transformation of the XG theory as follows:
\begin{align}
\label{ict3}
    &{\cal G} = {\cal G}_P + \int d^3x \left(\sum_a C_{\pi_a} + C\right) \ , \nonumber \\
    &C= \sqrt{h}\left( c_1(t) + c_2(t)R\right) \ , \nonumber  \\
    &\cph = c_3(t) \frac{\pi_h^2}{\sqrt{h}} + c_4(t)\frac{\pi_h^{ij}\pi_{hij}}{\sqrt{h}} \ , \nonumber \\
    &\cn = \cn[\pi_N, \pi_h^{ij}; N, h_{ij}, t] \ , \nonumber \\
    &\ci = \ci[\pi_i, \pi_h^{ij}; N, h_{ij}, t] \ , \nonumber \\
    &\ca = c_3\frac{\pi_h}{\sqrt{h}}\left( 2\pi_\Lambda^{ij}\Lambda_{ij} - \frac{1}{2}\pi_\Lambda\Lambda \right) + c_4\frac{\pi_{hij}}{\sqrt{h}} \left( 2\pi_\Lambda^{jk}\Lambda_{k}^i - \frac{1}{2}\pi_\Lambda^{ij}\Lambda \right)
     + c_5\left( \frac{\pi_h\pi_\Lambda^{ij}}{\sqrt{h}}\Lambda_{ij} - \frac{\pi_h^{ij}\pi_\Lambda}{\sqrt{h}}\Lambda_{ij}\right) \nonumber \\
     &\qquad+\frac{1}{2N}\left( \dot{c}_3\frac{\pi_h\pi_\Lambda}{\sqrt{h}} + \dot{c}_4\frac{\pi_{hij}\pi_\Lambda^{ij}}{\sqrt{h}}\right) -\frac{(\cn)_{\pi_N}}{N}\pi_\Lambda^{ij}\Lambda_{ij} - \frac{1}{N}\pi_\Lambda^{ij}D_i(\ci)_{\pi j}\ .
\end{align}
The total Hamiltonian $H_{tot}$ transforms as
\begin{align}
\label{htot}
    H_{tot}&= \bar{H}_{tot} + \epsilon ({\cal G})_t \nonumber \\
              &= \int d^3x \left[ \bN^i\bar{{\cal H}}_i + \bN \sqrt{\bh}\left( \bar{F}_{\bar{\Lambda}}^{ij}
               \bar{\Lambda}_{ij} - \bar{F}\right) \right] + \int d^3x\left( \bar{\lambda}_N\bp_N +
               \bar{\lambda}^i\bp_i + \bar{\lambda}_{\Lambda ij}\bp_\Lambda^{ij} + \bar{\lambda}_{hij}\tbp_h^{ij} \right) + \epsilon ({\cal G})_t \nonumber \\
    &= \int d^3x \left[ N^i\ch + N\sqrt{h}\left( F_\Lambda^{''ij}\Lambda_{ij} - F''\right)\right]
    + \int d^3x\left( \lambda_N\pi_N +
               \lambda^i\pi_i + {\lambda}_{\Lambda ij}\pi_\Lambda^{ij} + {\lambda}_{hij}\tilde{\pi}_h^{ij} \right) \ , \\
               \label{F5}
     F'' &\approx F - \epsilon \, \delta F'' \ , \nonumber \\
     \delta F'' &= \left\{ F_NP_N +aF_h +\left( a-\frac{P_N}{N}\right)F_\Lambda^{ij}\Lambda_{ij}
    +\frac{\dot{a}}{2N}F_\Lambda -\frac{1}{N}F_\Lambda^{ij}D_i\left(P_j + (\ci)_{\pi j}\right) \right. \nonumber \\
         &\quad+ \left( F_N - \frac{1}{N}F_\Lambda^{ij}\Lambda_{ij}\right)(\cn)_{\pi_N} +c_1\Lambda + \frac{\dot{c}_1}{N} -2c_2G^{ij}\Lambda_{ij} + \frac{\dot{c}_2}{N}R \nonumber \\
         &\quad+c_3F_\Lambda\left( F_h + F_\Lambda^{ij}\Lambda_{ij}-\frac{1}{4}F_\Lambda\Lambda\right)+\frac{\dot{c}_3}{4N}F_\Lambda^2
         +\left. c_4F_\Lambda^{ij}\left( F_{hij} + {F_\Lambda}_i^k\Lambda_{kj}-\frac{1}{4}F_{\Lambda ij}\Lambda \right)+\frac{\dot{c}_4}{4N}F_\Lambda^{ij}F_{\Lambda ij}\right\} \ ,
\end{align}
where $F_\Lambda\equiv h_{ij}F^{ij}_\Lambda$.
We note that the coefficient $c_5$ does not appear in the transformed total Hamiltonian (\ref{htot}), and thus we do not need to include it in the generator practically.
If we return to the Lagrangian formalism again by the Legendre transformation, we will obtain the transformed action which has the same form as (\ref{SA}) by replacing $F$ with $F''$.
We found the generator of the infinitesimal invariant canonical transformation as well as the invariant point transformation under the assumptions (\ref{as}) and (\ref{cph}).
In general, it is difficult to find the invariant canonical transformations of a theory. However,
we find such transformation by using the infinitesimal canonical transformations,
this helps us to understand the symmetry of the theory and the relations to other theories.

If we consider only $C$ as the generator, we can immediately promote the infinitesimal transformation to the finite transformation.
Removing the infinitesimal parameter $\epsilon$, we give the generator of the finite canonical transformation as
\begin{equation}
\label{ }
{\cal W} = \int d^3x \left\{ \pi_N\bN + \pi_i\bN^i + \pi_h^{ij}\bh_{ij} + \pi_\Lambda^{ij}\bar{\Lambda}_{ij}
             + \sqrt{\bh}\left(c_1 + c_2 \bar{R}\right)\right\} \ .
\end{equation}
Combining the above generator with Eq. (\ref{ct3}), we get
 \begin{align}
\label{ }
\bp_h^{ij} &= \pi_h^{ij} + \frac{1}{2}\sqrt{h}c_1h^{ij} - \sqrt{h}c_2G^{ij} \ , \nonumber \\
 H &= \bar{H} + \int d^3x \sqrt{h}\left(\dot{c}_1 + \dot{c}_2 R\right) \ .
\end{align}
Redefine the functional $\bar{F}$ to $F''$ in the same way as the infinitesimal transformation
without $\epsilon$,
\begin{equation}
\label{CCF}
F'' = F - \left( c_1\Lambda -2c_2G^{ij}\Lambda_{ij} +\frac{\dot{c}_1}{N}+\frac{\dot{c}_2}{N}R\right) \ .
\end{equation}
we obtain
\begin{align}
\label{}
    \tbp_h^{ij}& = \pi_h^{ij} - \frac{\sqrt{h}}{2}F_\Lambda^{''ij} \ , \nonumber  \\
    H& = \int d^3x \left[ -2N^iD_j\left( \frac{{\pi_h}_i^j}{\sqrt{h}} + \frac{1}{2}c_1\delta_i^j - c_2G_i^j\right)
            + N\sqrt{h}\left( F_\Lambda^{''ij}\Lambda_{ij} - F''\right)\right] \nonumber \\
       &=  \int d^3x \left[ N^i\ch
            + N\sqrt{h}\left( F_\Lambda^{''ij}\Lambda_{ij} - F''\right)\right] \ .
\end{align}
Therefore, in addition to the finite point transformation, the XG theory is invariant under the finite canonical transformation generated by $C$.

\section{Canonical connection of the Horndeski theory and the GLPV theory}\label{sec5}

The Horndeski theory and the GLPV theory in the unitary gauge are included in the 4- and 6-parameter models\cite{Gao:2014soa} of the XG theory.
We derived the infinitesimal invariant canonical transformation of the XG theory in the previous section. In this section,
using the invariant canonical transformation, we investigate the transformations which leave the Horndeski theory or the GLPV theory invariant,
and search the relation between the two theories in the unitary gauge.

We mentioned in Section \ref{sec1} that the Horndeski theory is invariant under the disformal transformation in which the conformal
and disformal factors depend only on the scalar field $\phi$ \cite{Bettoni:2013diz}. On the other hand,
the GLPV theory is invariant under the derivative dependent  disformal transformation in which
the disformal factor depends not only on the scalar field but also on the gradient of the scalar filed,
$X=-g^{\mu\nu}\nabla_\mu\phi\nabla_\nu\phi$ \cite{Gleyzes:2014qga, Crisostomi:2016tcp, Crisostomi:2016czh}.
Whilst the GLPV theory is regarded as more general scalar tensor theory than Horndeki theory
since the GLPV theory has two more terms than the Horndeski theory has, it was shown that a particular class of the GLPV theory
can be mapped into the Horndeski theory by the derivative
dependent disformal transformation \cite{Gleyzes:2014qga, Crisostomi:2016tcp, Crisostomi:2016czh}.
In the unitary gauge, the derivative dependent  disformal transformations  are included in the invariant
point transformation (\ref{lt}) we found in the previous section.

We derived the invariant canonical transformation of the XG theory which is larger than the point transformation.
Thus, if we use the invariant canonical transformation, we expect that we can find a larger class
of the invariant transformations of the Horndeski theory and the GLPV theory, and that
we may obtain general classes of the GLPV theory from the Horndeski theory.
In the following, firstly, we derive the invariant canonical transformations of the Horndeski theory and the GLPV theory using (\ref{ict3}).
Then we investigate the relation between the two theories under the canonical transformation and
focus on whether we can obtain general GLPV theory from the Horndeski theory or not.

First, we consider the invariant canonical transformation of the GLPV theory.
We begin with the GLPV theory in the barred system. In the unitary gauge, the functional $\bar{F}$ for the GLPV theory expressed with $\bar{\Lambda}_{ij}$ is given as \cite{Gleyzes:2014dya}
\begin{align}
\label{GLPV}
    \bar{F}&= \sum_{i=2}^5\bar{L}_i \ ,  \nonumber \\
    \bar{L}_2&= \bar{A}_2(t, \bN) \ , \nonumber \\
    \bar{L}_3&= \bar{A}_3(t, \bN)\bar{\Lambda} \ , \nonumber \\
    \bar{L}_4&= \bar{A}_4(t, \bN)\bar{\Lambda}_2 + \bar{B}_4(t, \bN)\bar{R} \ , \nonumber \\
    \bar{L}_5&= \bar{A}_5(t, \bN)\bar{\Lambda}_3 + \bar{B}_5(t, \bN)\bar{G}^{ij}\bar{\Lambda}_{ij} \ , \nonumber \\
    \bar{\Lambda}_2& \equiv \bar{\Lambda}^2 - [\bar{\Lambda}^2],\quad \bar{\Lambda}_3 \equiv \bar{\Lambda}^3 - 3\bar{\Lambda}[\bar{\Lambda}^2] + 2[\bar{\Lambda}^3] \ ,
\end{align}
where the trace products of a tensor $T_{ij}$ is defined as $[T^n]\equiv T^i_{j_1}T^{j_1}_{j_2}\cdots T^{j_{n-1}}_i$.
The GLPV theory has six independent `parameters', $(\bar{A}_i, \bar{B}_i)$. From $\bar{F}$,
we obtain the functional $F''$  from Eq. (\ref{F5}) by the infinitesimal canonical transformation (\ref{ict3}).
We restrict the generator of the canonical transformation so that $F''$  has the GLPV form (\ref{GLPV}) in the unbarred system.
The form of the functional $F$ in the unbarred system is the same as $\bar{F}$, and thus it preserves the GLPV form.
The first order terms of $\epsilon$ in $F''$, $\delta F''$, have the derivatives of $F$ and are derived in Appendix C
for the GLPV theory. From Eq. (\ref{FF5}), we find  that $\delta F''$ has at most the fifth powers of $\Lambda_{ij}$
\begin{align}
\label{ }
\delta F'' &\supset \left( (A_5)_N -\frac{2A_5}{N}\right)\Lambda_3 \cdot(\cn)_{\pi_N}\left[\frac{\sqrt{h}}{2}F_\Lambda^{ij}\right]
           + \frac{9}{2}c_3A_5^2\left( \Lambda_2\Lambda_3 -\frac{1}{2}\Lambda\Lambda_2^2\right) \nonumber \\
       &+ \frac{3}{4}c_4A_5^2\left\{ 2\Lambda_2\Lambda_3 + \Lambda\left( \Lambda_2^2 -4\Lambda^2[\Lambda^2] +8\Lambda[\Lambda^3] - 4[\Lambda^4]\right) \right\} \ ,
\end{align}
and we must eliminate these $\Lambda_{ij}$ terms since they are not included in the GLPV theory.
The fifth powers of $\Lambda_{ij}$ include the $\Lambda[\Lambda^4]$ term which cannot be cancelled with the other terms,
so we must set $c_4=0$ for the general coefficient $A_5$. Once we set $c_4=0$, the $\Lambda\Lambda_2^2$ term with the coefficient $c_3$ cannot be cancelled by the other terms,
thus we set $c_3=0$. To eliminate the remaining terms, we set $\cn =0$. As a result, we need $\cn,\,c_3,\,c_4=0$ to eliminate
the fifth powers of $\Lambda_{ij}$ in $\delta F''$.

From Eq. (\ref{FF4}), we get
\begin{align}
\label{FF3}
    \delta F''|_{\cn, c_3, c_4=0}& \approx -\frac{3}{N}A_5\left(\Lambda_2h^{ij} -2\Lambda\Lambda^{ij}+ 2\Lambda^{(i}_k\Lambda^{kj)}\right)D_i\left( P_j+(\ci)_{\pi j}\left[\frac{\sqrt{h}}{2}F_\Lambda^{ij}\right]\right)  \nonumber \\
      &\quad+ \left\{P_N\left( (A_5)_N - \frac{2A_{5}}{N}\right) +\frac{3}{2}aA_5\right\}\Lambda_3 \ .
\end{align}
The quartic terms of $\Lambda_{ij}$ come from the $\ci$ term only, thus we set $\ci=0$.
From the cubic terms of $\Lambda_{ij}$ in the second line of Eq. (\ref{FF3}),
we find that $P_N$ must depend on $N$ and $t$ only, i.e., $P_N = P_N[N,t]$,
to include the cubic terms in $L_5$ in the GLPV theory.
Note that the $P^i$ term contains the spatial covariant derivative. The GLPV theory does not contain such a term,
so we must eliminate the spatial derivative term. Since $P^i$ is independent of $\Lambda_{ij}$, we cannot
put the $P^i$ term to be a total derivative. Therefore, we set $P^i=0$.

In summary, the generator of the invariant canonical transformation of the GLPV theory are constructed by 
\begin{align}
\label{GLI}
&\cn = \ci = \cph= \ca= 0, \quad C = \sqrt{h}(c_1(t) + c_2(t)R), \nonumber \\
 &P_N=P_N[N,t],\quad P^i=0,\quad P_{ij}= a(t)h_{ij}, \quad P_{\Lambda ij} = \left( a -\frac{P_N}{N}\right) \Lambda_{ij} + \frac{\dot{a}}{2N}h_{ij}\ .
\end{align}
With them, we get
\begin{align}
\label{ }
 \int d^3x N\sqrt{h}F'' &=\int d^3x\sqrt{h}\left[ NF- \epsilon\left\{ \left( NF_N - F_\Lambda^{ij}\Lambda_{ij}\right)P_N + a\left(NF_h +NF_\Lambda^{ij}\Lambda_{ij}\right)+ \frac{\dot{a}}{2}F_\Lambda  \right.\right. \nonumber  \\
     &\qquad\qquad\qquad +\dot{c}_1+Nc_1\Lambda +\dot{c}_2R -2Nc_2G^{ij}\Lambda_{ij}\bigg\}  \bigg] \nonumber \\
     &= \int d^3xN\sqrt{h}\left[ A_2''+A_3''\Lambda+A_4''\Lambda_2+A_5''\Lambda_3 +B_4''R+B_5''G^{ij}\Lambda_{ij}\right] \ , \nonumber \\
     A_2''&= A_2 -\epsilon\left[ \left( (A_2)_N +\frac{A_2}{N}\right)P_N +\frac{3}{2}aA_2
     + \frac{3\dot{a}}{2N}A_3+\frac{\dot{c}_1}{N}\right]  \ , \nonumber \\
     A_3''&= A_3 -\epsilon\left[ (A_3)_N P_N +\frac{3}{2}aA_3 +\frac{2\dot{a}}{N}A_4 + c_1\right] \ , \nonumber   \\
     A_4''&= A_4 -\epsilon\left[ \left( (A_4)_N -\frac{A_4}{N}\right)P_N +\frac{3}{2}aA_4
    +\frac{3\dot{a}}{2N}A_5\right] \ , \nonumber    \\
    A_5''&= A_5 -\epsilon\left[ \left( (A_5)_N -\frac{2A_5}{N}\right)P_N +\frac{3}{2}aA_5\right] \ , \nonumber    \\
     B_4''&= B_4 -\epsilon\left[ \left( (B_4)_N +\frac{B_4}{N}\right)P_N +\frac{1}{2}aB_4
    -\frac{\dot{a}}{4N}B_5 + \frac{\dot{c}_2}{N}\right] \ , \nonumber    \\
     B_5''&= B_5 -\epsilon\left[ (B_5)_NP_N +\frac{1}{2}aB_5 -2c_2\right]  \ .
\end{align}
Note that all the covariant derivative terms created by $F_h$ in $\delta F''$ vanish as the total derivatives. The coefficients $(A_i,\,B_i)$ have the same functional forms as $(\bar{A}_i,\,\bar{B}_i)$ with unbarred arguments. The transformed GLPV theory has the coefficients $(A''_i,\, B''_i)$ constructed from $(A_i,\,B_i)$ and $(c_1,\,c_2,\,a,\,P_N)$. 

Now we can directly promote the above infinitesimal invariant transformation to the finite transformation.
According to the argument in Section \ref{sub4}, we can promote the infinitesimal point transformation
to the finite transformation by removing the infinitesimal parameter $\epsilon$. When $P_N=P_N[N,t]$, $P^i=0$ and $a=a(t)$,
the finite invariant point transformation of the GLPV theory is
\begin{align}
\label{gi}
    &\bar{N}=\sqrt{{\cal A}N^2 -{\cal B}},\quad \bar{N}^i=N^i,\quad \bh_{ij} = {\cal A}(t)h_{ij}  \ , \nonumber \\
    &{\cal B}[t, N] = b = {\cal A}N^2 -(N-P_N[N, t])^2  \ .
\end{align}
Restoring the time diffeomorphism by the scalar field $\phi$, we obtain the general covariant
form of the finite invariant point transformation of the GLPV theory as
\begin{align}
\label{GL2}
    &\bar{g}_{\mu\nu}= {\cal A}[\phi]g_{\mu\nu} +{\cal B}[\phi, X]\nabla_\mu\phi\nabla_\nu\phi    \ , \nonumber \\
    &{\cal B}[\phi, X]= b = \frac{{\cal A}}{X} - (X^{-1/2}-{\cal P}_N[\phi, X])^2 \ .
\end{align}
Thus, the invariant point transformation derived from the infinitesimal transformation corresponds to the familiar general disformal transformation
which leaves the GLPV theory invariant. Moreover, we can add another invariant transformation generated by $C$.
As derived in Section \ref{sub43}, the finite canonical transformation generated by $C$ gives the functional $F''$ in Eq. (\ref{CCF}),
and all the terms generated by $C$ belong to the GLPV theory. Therefore, in addition to the general disformal transformation \eqref{GL2},
we found additional canonical transformation generated by $C$ which leaves the GLPV theory invariant.

As a special case, we consider the infinitesimal transformation which leaves the Horndeski theory invariant.
The functional $\bar{F}$ of the Horndeski theory has the same form as that of the GLPV theory,
but the coefficients must satisfy the two conditions,
\begin{align}
\label{HC}
    \bar{A}_4&= -\bar{B}_4 +2\bar{X}(\bar{B}_4)_{\bar{X}}\  \rightarrow\ \bar{A}_4= -\bar{B}_4 -\bar{N}{(\bar{B}_4)}_{\bar{N}} \quad {\rm in\ the\ unitary\ gauge}  \ , \nonumber \\
         \bar{A}_5 &= -\frac{\bar{X}{(\bar{B}_{5})}_{\bar{X}}}{3}
   \quad\qquad \rightarrow\ \bar{A}_5 = \frac{\bar{N}(\bar{B}_{5})_{\bar{N}}}{6}
   \quad {\rm in\ the\ unitary\ gauge}\ .
\end{align}
Thus, the Horndeski theory has two less independent coefficients than the GLPV theory does.
The coefficients $(A_4,\,B_4)$ and $(A_5,\,B_5)$ in the unbarred system also satisfy these conditions since the coefficients $(A_i,\,B_i)$ have the same forms as $(\bar{A}_i,\,\bar{B}_i)$. In order to make the unbarred system be the Horndeski theory, we need to constrain $(c_1, c_2, a, P_N)$ so that the coefficients $(A_4'',\,B_4'')$ and $(A_5'',\,B_5'')$ satisfy the Horndeski conditions (\ref{HC}).
We use the functional $b[N, t]$ instead of $P_N$ to compare the result in the general covariant form.
Under the infinitesimal transformation, using Eq. (\ref{gi}), $b$ becomes
\begin{align}
\label{}
    b=(1-\epsilon a)N^2 -(N-\epsilon P_N)^2 = \epsilon\left( - aN^2+2NP_N \right) + {\cal O}(\epsilon^2) \ .
\end{align}
Redefining $b$ to $\epsilon b$, $P_N$ is expressed as
\begin{equation}
    P_N = \frac{a}{2}N + \frac{b}{2N} \ .
\end{equation}
Using this relation, we obtain
\begin{align}
\label{thc}
    &A_4''= 
    A_4 -\epsilon\left[ \left( (A_4)_N \right)P_N +\left(a-\frac{b}{2N^2}\right)A_4
    +\frac{3\dot{a}}{2N}A_5\right] \ ,\nonumber \\
     -&B''_4-N(B''_4)_N= -B_4 -N(B_4)_N + \epsilon\bigg[ \left(2(B_4)_N+N(B_4)_{NN}\right)P_N  \nonumber \\
           &\qquad\qquad\qquad\qquad\quad+\left.\left(a-\frac{b}{2N^2}\right)(B_4+N(B_4)_N)- \frac{\dot{a}}{4}(B_5)_N
              +\left(\frac{B_4}{2N}+\frac{(B_4)_N}{2}\right) (b)_N\right] \ , \nonumber \\
     &A''_5 = A_5 -\epsilon\left[ (A_5)_NP_N + \left( \frac{1}{2}a -\frac{b}{N^2}\right)A_5\right] \ ,
     \nonumber \\
     &\frac{N(B''_5)_N}{6} = \frac{N(B_5)_N}{6} -\epsilon\left[  \frac{1}{6}\left((B_5)_N
             +N(B_5)_{NN}\right)P_N + \left( \frac{1}{2}a -\frac{b}{N^2}\right)\frac{N(B_5)_N}{6}
             +\frac{(B_5)_N}{12}(b)_N \right] \ ,
\end{align}
where $(B_i)_{NN}\equiv \int d^3x'\delta (B_i)_N(\vxx)/\delta N$. The relations between
$(A_4'',\,B_4'')$ and $(A_5'',\,B_5'')$ become
\begin{align}
\label{HC2}
    A''_4& = -B''_4 -N(B''_4)_N -\epsilon\left(\frac{B''_4}{2N}+\frac{(B''_4)_N}{2}\right) (b)_N \ , \nonumber \\
    A''_5&=  \frac{N(B''_5)_N}{6} + \epsilon \frac{(B''_5)_N}{12}(b)_N \ .
\end{align}
We change the coefficients $B_i$ to $B_i''$ in the first order terms of the $\epsilon$ in Eq.(\ref{HC2})
since that procedure creates up to second order terms of $\epsilon$.
If the functional $b$ is independent of $N$, i,e,, $(b)_N=0$, we obtain the Horndeski
conditions for $(A_4'',\,B_4'')$ and $(A_5'',\,B_5'')$ also, thus the transformed theory is the Horndeski theory with the coefficients $(A''_i,\,B''_i)$. Note that the deviations from the Horndeski conditions do not depend on $c_1(t)$ and $c_2(t)$,
thus the $N$ dependence of $b$ completely determines whether the transformed theory is the Horndeski theory or not.
If we promote this result to the general covariant form, we find that the Horndeski theory is invariant under the disformal transformation
\begin{equation}
\label{}
   \bar{g}_{\mu\nu}= {\cal A}[\phi]g_{\mu\nu} +{\cal B}[\phi]\nabla_\mu\phi\nabla_\nu\phi   \ .
\end{equation}
Thus, the invariant point transformation obtained from the infinitesimal transformation corresponds to the familiar disformal transformation
which leaves the Horndeski theory invariant. As in the GLPV theory, we can add the invariant canonical transformation generated by $C$ also for the Horndeski theory since the Horndeski conditions (\ref{HC}) are still satisfied if we add the transformation generated by $C$.

Can we transform the Horndeski theory in the barred system to a general GLPV theory in the unbarred system
by the canonical transformation we found? The answer is no.
In order to transform the Horndeski theory to a general GLPV theory, we need not only to
break the Horndeski conditions (\ref{HC}) but also to create two more independent functionals of $N$ and $t$
than the Horndeski theory has. Starting from the Horndeski theory,
the Horndeski conditions transform as (\ref{HC2}). If the functional $b$ depends on $N$,
then the transformed coefficients $(A''_4,\,B''_4)$ and $(A''_5,\,B''_5)$ violate the Horndeski conditions,
and the transformed theory becomes a particular class of the GLPV theory.
However, we can deviate $A''_4$ and $A''_5$ from the Horndeski conditions only by the amount generated by $(b)_N$.
This implies that if we determine the transformation of $A''_4$ with $b$, the transformation of $A''_5$ is also determined at the same time,
thus $A''_4$ and $A''_5$ are not totally independent from each other. This situation does not improve even if we include the canonical
transformation generated by $C$ since the deviations from the Horndeski conditions are independent of $C$.
The same conclusion is obtained even for the finite transformation because the finite transformation is obtained by repeating
the infinitesimal transformation, so $A''_4$ and $A''_5$ are not independent. Therefore, we can not transform the Horndeski
theory to  general GLPV theory through the canonical transformation we found in this article.

\section{Summary}\label{sec6}

We studied the number of degrees of freedom and the invariant canonical transformation of the XG theory.
Using the invariant transformation, we studied the relation between the Horndeski and GLPV theories
which are special cases of the XG theory in the unitary gauge.

After we briefly reviewed the XG theory, we performed the full Hamiltonian analysis
of the XG theory to elucidate the number of degrees of freedom of the theory.
The Hamiltonian analysis for the XG theory was performed before only in a perturbative way.
The perturbative analysis may break down when the interactions become strong compared to the quadratic terms in the theory.
To overcome this, we introduced an auxiliary field and rewritten the action of the XG theory
so that we can perform the full Hamiltonian analysis without using perturbations.
As a result of the Hamiltonian analysis, we confirmed that the number of degrees of freedom for the XG theory is no more than 3,
which corresponds to the two tensor modes plus a scalar mode in (3+1) dimensions,
as long as the theory have the spatial diffeomorphism. Therefore,
the XG theory is free from any extra degree of freedom which causes the Ostrogradski instability,
and the theory is stable.

Under the infinitesimal transformation, the variations generated by the transformation are treated as
infinitesimal quantities since they are multiplied by the infinitesimal parameter $\epsilon$,
so we keep the first order terms of $\epsilon$ only.
The infinitesimal invariant point transformation was first discussed in the Lagrangian formalism,
then it was  promoted to finite transformation by repeating the infinitesimal transformation.
If we restore the time diffeomorphism symmetry by introducing
the Stuck\"elberg scalar field $\phi$ and express the quantities in the general covariant form,
the invariant point transformation becomes the metric transformation,
and it includes not only the familiar disformal transformation but also the transformation generated by the vector product.
The disformal factor ${\cal B}$ and the vector product $V^{\mu}$ could include higher order derivative terms as
long as they consist of the building blocks of the XG theory in the unitary gauge $\phi=t$.
Based on the arguments in \cite{Langlois:2015cwa, Langlois:2015skt, Crisostomi:2016tcp,Crisostomi:2016czh}, however,
it is uncertain whether all the classes of the general covariant form of the XG theory become stable,
while the spatially covariant form of the XG theory is stable as we proved in Section \ref{sec3}.
Thus we should be careful with the invariant metric transformation in the general covariant form.
The possibility that the metric transformation may include some special classes of higher order derivative terms while
keeping the theory the stability will provide another direction for searching the equivalence between different theories
and looking for the symmetries of theories.

The metric transformations considered before are included in the point transformation in the unitary gauge.
We extended the metric transformation to the canonical transformation. 
We derived the canonical transformation which leaves the XG theory invariant.
We applied the infinitesimal canonical transformation in the Hamiltonian formalism to find the invariant canonical transformation.
Under a few assumptions, we found the infinitesimal invariant canonical transformation of the XG theory
which includes the point transformation. We also promoted the infinitesimal transformation generated by a part of generator $C$
to finite transformation by removing the infinitesimal parameter $\epsilon$.

Using the infinitesimal invariant canonical transformation of the XG theory, we obtained the
transformations which leave the GLPV theory and the Horndeski theory invariant, and studied the relation
between the two theories in the unitary gauge. It has been known that both theories are invariant under
the disformal transformations. We derived the infinitesimal transformations of the Hamiltonian for both theories,
and discovered that the both theories are invariant not only under the disformal transformations
but also the additional canonical transformation generated by $C$.
We also investigated whether we can obtain general GLPV theory from the Horndeski theory
through the invariant canonical transformation we found.
Whilst the coefficients $\bar{A}_4$ and $\bar{A}_5$ in the Horndeski theory must satisfy the Horndeski conditions (\ref{HC}),
those coefficients in GLPV theory do not need to satisfy the conditions and we are free to choose
the coefficients $A''_4$ and $A''_5$ in the GLPV theory.
In order to obtain the general GLPV theory from the Horndeski theory,
we need to deviate the two coefficients $\bar{A}_4$ and $\bar{A}_5$ in the Horndeski theory independently from the Horndeski conditions.
It is well known that only a particular class of GLPV theory can be obtained from the Horndeski theory via the disformal transformation.
We confirmed that even if we consider the canonical transformation generated by $C$ in addition to the disformal transformation,
we can not obtain the coefficients $A''_4$ and $A''_5$ in the GLPV theory independently from the Horndeski theory,
so we can not transform the Horndeski theory to a general GLPV theory through the canonical transformation we found in this article.
Moreover, the generator $C$ does not affect the violations of the Horndeski conditions (\ref{HC}) for $A''_4$ and $A''_5$.
Although we include the additional canonical transformation generated by $C$, we can not obtain more general GLPV theory
than the particular class obtained from the Horndeski theory by the disformal transformation.

On the other hand, as shown in Section \ref{sec5}, the invariant canonical transformation can create higher power of $\Lambda_{ij}$ or $K_{ij}$ which any point transformations can not create. Therefore, for example, we can connect the Horndeski theory to another scalar-tensor theory which has higher power of $\Lambda_{ij}$ or $K_{ij}$ than the Horndeski theory by using the canonical transformation. 

In this article, we focused on the XG theory only, and we derived the invariant transformation of the XG theory using the infinitesimal canonical transformation.
The procedure  we provided to study the relation between two theories, of course, can apply not only to the XG theory but also to any field theory. Therefore, the procedure is a useful method to study the equivalence between two different theories and the symmetries of a theory.


\section*{\bf Acknowledgments}
The author greatly appreciate Y. Gong modifying the manuscript of this article.
This work is partially supported by the Natural Science Foundation of
China under Grant No. 11475065.

\appendix
\section{The Hamiltonian analysis for the special cases}

\subsection{The case the Green's function $G^{xy}_{\Lambda\Lambda, ij,kl}$ does not exist}\label{app1}
In this subsection, we consider the case when $G_{\Lambda\Lambda,ij,kl}^{xy}$ does not exist for the Hamiltonian analysis in Section \ref{sec3}. We have the primary constraints as
\begin{equation}
\label{ }
\pi_N \approx0,\quad \pi_i \approx0,
\quad \pa\approx0,\quad
\tilde{\pi}_h^{ij}\equiv \pi_h^{ij} - \frac{\sqrt{h}}{2}F_\Lambda^{ij} \approx 0 \ .
\end{equation}
In Eq. \eqref{pineq1}, we define the linear combination $\tilde{\pi}_N$ of the constraints by using $G_{\Lambda\Lambda,ij,kl}^{xy}$.
However, when $G_{\Lambda\Lambda,ij,kl}^{xy}$ does not exist, such a linear combination cannot be defined, and thus
it is unclear whether its consistency condition creates a secondary constraint such as $\chi$.
On the other hand, the Poisson brackets of the constraint $\pi_i$ with all primary constraints
vanish even if $G_{\Lambda\Lambda,ij,kl}^{xy}$ does not exist, and thus its consistency condition creates the secondary constraint $\ch$
even if $G_{\Lambda\Lambda,ij,kl}^{xy}$ does exist.

The XG theory has the spatial diffeomorphism symmetry by construction so that the theory must have the
first class constraints which consist of the generator of the spatial diffeomorphism group.
The variation of physical quantities $Y(\vx)$ under the gauge transformation for the classical theory is given as
\begin{align}
\label{ge}
\delta Y(\vx) &= \left\{ Y(\vx) , {\cal G}[\theta^\alpha]\right\}_P \ , \nonumber \\
 {\cal G}[\theta^\alpha] &= \sum_\alpha\int d^3x'\theta^\alpha\Psi_\alpha \ ,
\end{align}
where ${\cal G}$ is the generator of the gauge transformation, $\theta^\alpha$ are the independent gauge parameters and $\Psi_\alpha$
are the independent first class constraints.
Considering the variation of shift vector under the spatial diffeomorphism in Eq. (\ref{va}),
we find that the variation has 6 independent gauge parameters, $\eta^i$ and $\xi^i$. $\eta^i$ is given
as the time derivative of $\xi^i$. In general, the time derivative of a spatial vector is independent
of the original vector on the constant time hypersurface. Thus, $\eta^i$ and $\xi^i$ are independent of each other.
We need 6 gauge generators  of the spatial diffeomorphism whose gauge orbits are parametrized by $\eta^i$ and $\xi^i$ to
construct the generator ${\cal G}$. In order to derive the variation given in the Eq. (\ref{va})
as well as $\delta \Lambda_{ij} = {\frak L}_\xi \Lambda_{ij}$ through the relation (\ref{ge}), we must have the following generator for the spatial diffeomorphism:
\begin{equation}
\label{ }
\int d^3x'\left[ \eta^i\pi_i + \xi^i\mom\right] \ .
\end{equation}
Therefore, $\pi_i$ and $\mom$ are the first class constraints even if $G_{\Lambda\Lambda,ij,kl}^{xy}$ does not exist. Up to this stage, we have the 6 first class constraints, $\pi_i$ and $\mom$, and the  13 constraints, $\pi_N$, $\pi_\Lambda^{ij}$ and $\tilde{\pi}_h^{ij}$ at least. The number of degrees of freedom becomes
\begin{equation}
\label{ }
\# \leq \frac{1}{2}(32-2\times6-13) = 3.5 \ .
\end{equation}
The 0.5 degree of freedom should not exist, and thus we can ensure that the XG theory has up to 3 degrees of freedom no matter whether the Green's function $G_{\Lambda\Lambda,ij,kl}^{xy}$ exists or not.

\subsection{The case the XG theory has the general covariance}
In this subsection, we investigate the structure of constraints for the general covariant class of XG theory, which has the general covariance even if we do not restore the symmetry under the time diffeomorphism by the scalar field $\phi$. If a class of the XG theory has the general covariance, that class should have exactly the same structure of constraints and Poisson algebra. Therefore, just by studying a certain general covariant model, we can know the structure of constraints for all models of the general covariant class of the theory. In this subsection, we consider only the class in which the Green's function $G_{\Lambda\Lambda}^{xy,ij,kl}$ exists.

General relativity is the most familiar general covariant theory of gravitation,
and the XG theory includes it by construction. In the case of general relativity,
the Green's function $G_{\Lambda\Lambda}^{xy,ij,kl}$ is proportional to the inverse of DeWitt metric.
In the following, we study the structure of the constraints of general relativity with the Hamiltonian (\ref{H}).
We then generalize the result of general relativity and derive the conditions of constraints with which the general covariant class of XG theory should be satisfied.

The Hamiltonian of general relativity in the form of (\ref{H}) is given as
\begin{equation}
\label{ }
H_{EG} = \int d^3x\left[N^i\ch + \frac{M_{{\rm Pl}}^2N\sqrt{h}}{2}(\Lambda_{ij}\Lambda^{ij} - \Lambda^2 -R)\right]\ ,
\end{equation}
where $M_{{\rm Pl}}$ is the Planck mass. The first order derivative of $F$ by $\Lambda_{ij}$ is obtained as
\begin{equation}
\label{ }
F_\Lambda^{ij} = M_{\rm Pl}^2(\Lambda^{ij}-\Lambda h^{ij}) \ ,
\end{equation}
and thus we find
\begin{align}
\label{}
    \fan^{x,ij}&= 0 \ , \nonumber  \\
    \faa^{x,ij,kl}& = M_{\rm Pl}^2\sqrt{h}\left\{\frac{1}{2}(h^{ik}h^{jl}+h^{il}h^{jk}) - h^{ij}h^{kl}\right\} \ , \nonumber \\
    G_{\Lambda\Lambda,ij,kl}^{xy} &=  \frac{1}{M_{\rm Pl}^2\sqrt{h}}\left\{\frac{1}{2}(h_{ik}h_{jl}+h_{il}h_{jk}) - \frac{1}{2}h_{ij}h_{kl}\right\}\dxy \ .
\end{align}
In the case of general relativity, $\tilde{\pi}_N$ is equal to $\pi_N$ since $\fan^{x,ij}=0$.
As for the spatially covariant case, we obtain the secondary constraints from the consistency condition of $\pi_N$ and $\pi_i$ as
\begin{align}
\label{}
    \dot{\pi}_N&= -\frac{\delta H_{EG}}{\delta N} = \frac{M_{{\rm Pl}}^2\sqrt{h}}{2}(\Lambda_{ij}\Lambda^{ij} - \Lambda^2 -R)= \chi   \ , \nonumber \\
    \dot{\pi}_i& = -\ch \ .
\end{align}
We perform the Hamiltonian analysis in the same way as in the main text. However,
unlike the spatially covariant case,  the Poisson bracket between $\tc$ and $\pi_N$ weakly vanishes since the constraint $\chi$ does not depend on the lapse function
\begin{equation}
\label{ }
\{\tc,\,\pi_N\}_P \approx \frac{\delta\chi}{\delta N}= 0 \ .
\end{equation}
$\chi$ does not depend on the time coordinate explicitly, therefore the consistency condition of $\tc$
can not determine the Lagrange multiplier of $\pi_N$ and does not create any new constraint
\begin{align}
\label{ }
\frac{d}{dt}\left< \tc,f\right> &\approx \left\{\left< \tc,f\right>,\, H\right\}_P  \nonumber \\
 &= \left\{\left< \tc,f\right>,\, \left< \ch,N^i\right>\right\}_P
    + \left\{\left< \tc,f\right>,\, \left< \chi,N\right>\right\}_P  \nonumber  \\
    &\approx  -\int d^3x \frac{\delta\left< \chi,f\right>}{\delta N}N^iD_iN
     + \left\{\left< \tc,f\right>,\, \left< \tc,N\right>\right\}_P \nonumber \\
     &\quad - \left\{\left< \tc,f\right>,\, \left<\hat{\chi}_\Lambda^{x,ij}\left[ \left< 2G_{\Lambda\Lambda,kl,ij}^{xx'}\,,\,\tilde{\pi}^{kl}(\vxx)\right>\right],N\right> \right\}_P
     + \left\{\left< \tc,f\right>,\, \left< \hat{\psi}^{x,ij}_h\left[\left<2G_{\Lambda\Lambda,ij,kl}^{x'x}\, ,\,\pi_\Lambda^{kl}(\vx)\right>
     \right],N\right>\right\}_P \nonumber \\
     &\approx 0 \ .
\end{align}
As expected, $\pi_N$ and $\tc$ become the first class constraints in general relativity which indicates
that the theory has the time diffeomorphism, and this structure should hold for
the general covariant class of theory in the framework of (\ref{H}).
Generalizing these results of general relativity, all models of the general covariant class of XG theory
with  $G_{\Lambda\Lambda,ij,kl}^{xy}$ should satisfy the conditions
\begin{equation}
\label{}
    \{\tc,\,\tilde{\pi}_N\}_P \approx0, \quad \frac{d}{dt}\tc \approx 0 \ ,
\end{equation}
in addition to the other relations between the constraints in the main text.
The general covariant class of XG theory has 8 first class constraints,
$\tilde{\pi}_N$, $\tc$, $\pi_i$ and $\mom$, and 12 second class constraints, $\pi_\Lambda^{ij}$ and $\tilde{\pi}_h^{ij}$. Thus,
the number of degrees of freedom for such class is
\begin{equation}
\label{ }
\frac{1}{2}(32-2\times 8 -12)= 2 \ ,
\end{equation}
which corresponds to the number of tensor modes in 3+1 dimension, and there appears no scalar degree of freedom.

\section{The derivations of $C$ and $C_{\pi_\Lambda}$}

We derive the functional $C$ first. As mentioned in Section \ref{sub43},
the invariant point transformation already leaves the XG theory invariant, so we set ${\cal G}_P=0$ in this appendix.
When ${\cal G}_P=0$, the Hamiltonian transforms as
\begin{align}
\label{}
    H &= \int d^3x\left[ N^i{\cal H}_i + N\sqrt{h}(F_\Lambda^{'ij}\Lambda_{ij}-F') + \delta {\cal H}\right] \ , \nonumber  \\
    \delta {\cal H} &=
    2\left\{ \left( {(\cph)_h}^i_j + {(C)_h}^i_j- \pi_h^{ik}(\cph)_{\pi_hjk}\right) D_iN^j \right. \nonumber \\
   &\quad\qquad\left.-{\pi_h}^i_jD_i\left((C_\pi)_\pi^j\right) -\frac{1}{2}\pi_h^{ik}N^jD_j(\cph)_{\pi_hik} \right\}  \nonumber \\
   &\quad+  (C_{\pi_h})_t+(C)_t  -N\sqrt{h}F_\Lambda^{'ij} (C_{\pi_\Lambda})_{\pi_\Lambda ij} \nonumber \\
   &\quad +\Lambda_{ij}\int  d^3x'd^3y\frac{\delta (\ca)_{\pi_\Lambda kl}(\vy)}
                      {\delta \Lambda_{ij}} \frac{\delta (N\sqrt{h}F')(\vxx)}{\delta \Lambda_{kl}(\vy)}\ .
\end{align}
In $\delta {\cal H}$,
only the two terms, $( C)_t$ and $2{(C)_h}^i_jD_iN^j$ do not depend on $\pi_h^{ij}$ and $\Lambda_{ij}$, so they will not induce any primary constraints.
We must eliminate the latter term to leave the theory invariant since it depends on $N^i$. We can eliminate it as  a total derivative by imposing
\begin{align}
\label{}
    {(C)_h}^i_j&\propto \sqrt{h}\delta^i_j,\ \sqrt{h}G^i_j \ , \nonumber  \\
    {(C)_h}^i_jD_iN^j &\propto \sqrt{h}D_iN^i,\  \sqrt{h}D_i(G^i_jN^j) \ ,
\end{align}
where $G^i_j$ is the spatial Einstein tensor.
Thus, the functional $C$ is completely determined as
\begin{equation}
\label{C}
C= \sqrt{h}\left(c_1(t) + c_2(t)R\right)  \ .
\end{equation}
So
\begin{align}
\label{ }
(C)_h^{ij} &= \sqrt{h}\left( \frac{c_1}{2}h^{ij} - c_2G^{ij}\right) \ , \nonumber \\
(C)_t &= \sqrt{h}\left( \dot{c}_1+ \dot{c}_2R\right) \ . \nonumber
\end{align}
These terms can not combine with other terms to become any primary constraint, so absorb them by the redefinition of the functional $F'$. We redefine the functional $F'$ as
\begin{equation}
\label{ }
F' \ \rightarrow \ F'' = F' -\epsilon\left( c_1\Lambda - 2c_2G^{ij}\Lambda_{ij} +\frac{\dot{c_1}}{N} +\frac{\dot{c_2}}{N}R\right) \ ,
\end{equation}
and using $F''$, we obtain
\begin{align}
\label{}
    &F_\Lambda^{'ij}\Lambda_{ij} -F' +\epsilon (C)_t = F_\Lambda^{''ij}\Lambda_{ij} - F'' \ , \nonumber\\
    \label{tbp}
    &\tbp_h^{ij} \approx \pi_h^{ij} - \frac{\sqrt{h}}{2}F_\Lambda^{''ij}
     +\epsilon \left\{ (\cph)_h^{ij}  -\frac{(\cn )_{\pi_N}}{2N}\sqrt{h}F_\Lambda^{''ij} \right. \nonumber \\
                    &\qquad \left.-\frac{1}{2N}\int  d^3x'd^3y\frac{\delta (\ca)_{\pi_\Lambda kl}(\vy)}
                      {\delta \Lambda_{ij}} \frac{\delta (N\sqrt{h}F'')(\vxx)}{\delta \Lambda_{kl}(\vy)}\right\} \ ,
\end{align}
where we set ${\cal G}_P=0$. 
In (\ref{tbp}), we rewrote the functional $F'$ to $F''$ in the first order terms of $\epsilon$
since this creates only up to second order terms of $\epsilon$ and we can neglect them in the infinitesimal transformation.
We expect that the constraint $\tbp_h^{ij}$ transforms to $\tilde{\pi}_h^{ij}$ by replacing $\bar{F}$ with $F''$, i.e.,
\begin{equation}
\label{eq}
\tbp_h^{ij} \rightarrow \tilde{\pi}_h^{ij} = \pi_h^{ij} - \frac{\sqrt{h}}{2}F_\Lambda^{''ij} \approx 0 \ .
\end{equation}
Now we determine $C_{\pi_a}$ so that Eq. (\ref{eq}) is satisfied.

If we choose $C_{\pi_h}$ by hand as
\begin{equation}
\label{cphh}
\cph = c_3(t) \frac{\pi_h^2}{\sqrt{h}} + c_4(t)\frac{\pi_h^{ij}\pi_{hij}}{\sqrt{h}} \ ,
\end{equation}
then the constraint (\ref{tbp}) transforms as
\begin{align}
\label{t2}
    \tbp_h^{ij} \approx \pi_h^{ij}- \frac{\sqrt{h}}{2}F_\Lambda^{''ij} + &\epsilon\left\{ 2c_3\frac{\pi_h\pi_h^{ij}}{\sqrt{h}} + 2c_4\frac{{\pi_h}^i_k\pi_h^{kj}}{\sqrt{h}} -\frac{(c_3\pi_h^2 + c_4\pi_{hkl}\pi^{kl}_h)}{2\sqrt{h}}h^{ij} \right. \nonumber   \\
    &\left.-\frac{(C_{\pi_N})_{\pi_N}}{N}\frac{\sqrt{h}}{2}F_\Lambda^{''ij} -\frac{1}{2N}\int  d^3x'd^3y\frac{\delta (\ca)_{\pi_\Lambda kl}(\vy)}
                      {\delta \Lambda_{ij}} \frac{\delta (N\sqrt{h}F'')(\vxx)}{\delta \Lambda_{kl}(\vy)}\right\} \ .
\end{align}
To obtain the weak equality (\ref{eq}), the first order term of $\epsilon$ in Eq. (\ref{t2}) should vanish at least weakly.
The second and the third terms in the first order terms of $\epsilon$  in (\ref{t2})  must be eliminated
by the $\ca$ term because of the structure of the indices. When $\ca$ includes the following terms
%
\begin{equation}
\label{ca1}
        \ca\supset  2c_4\frac{{\pi_h}_{i}^k\pi_\Lambda^{ij}\Lambda_{jk}}{\sqrt{h}} -\frac{1}{2}\left( c_3\frac{\pi_h\pi_\Lambda}{\sqrt{h}}\Lambda +c_4\frac{{\pi_h}_{ij}\pi_\Lambda^{ij}}{\sqrt{h}}\Lambda \right) \ ,
\end{equation}
in the first order terms of $\epsilon$ in (\ref{t2}), the second and the third terms are combined with the $\ca$ term to become the terms
proportional to Eq. (\ref{eq}):
\begin{equation}
\label{ }
2c_4\frac{{\pi_h}^i_k}{\sqrt{h}}\left( \pi_h^{kj} - \frac{\sqrt{h}}{2}F_\Lambda^{''kj}\right)
-\frac{h^{ij}}{2}\left( c_3\frac{\pi_h}{\sqrt{h}}h_{kl} + c_4\frac{{\pi_h}_{kl}}{\sqrt{h}}\right)
\left( \pi_h^{kl} - \frac{\sqrt{h}}{2}F_\Lambda^{''kl}\right) \ .
\end{equation}
Therefore, if the weak equality (\ref{eq}) is satisfied, these terms vanish weakly.
To eliminate the remaining terms in the same manner, $\ca$ should include terms like
%
\begin{align}
\label{ca2}
     \ca&\supset  \left\{(2c_3 + c_5)\frac{{\pi_h}}{\sqrt{h}} -\frac{(\cn)_{\pi_N}}{N}\right\} \pi_\Lambda^{ij}\Lambda_{ij} - c_5\frac{\pi_\Lambda\pi_h^{ij}\Lambda_{ij}}{\sqrt{h}} \ .
\end{align}
In addition to these terms, we need a few more terms to eliminate the extra terms in $\delta {\cal H}$.
Using Eqs.(\ref{C}) and (\ref{cphh}), we find the extra terms in $\delta {\cal H}$ which can not be eliminated by (\ref{ca1}) and (\ref{ca2}) are
\begin{equation}
\label{ }
\delta {\cal H} \supset -2\pi_h^{ij}D_i(\ci)_{\pi j} + \dot{c}_3\frac{\pi_h^2}{\sqrt{h}} + \dot{c}_4\frac{\pi_h^{ij}{\pi_h}_{ij}}{\sqrt{h}}
   -N\sqrt{h}F_\Lambda^{''ij} (C_{\pi_\Lambda})_{\pi_\Lambda ij} \ .
\end{equation}
Therefore, in order to eliminate them, we need to include the following terms in $\ca$
\begin{equation}
\label{ca3}
\ca\supset -\frac{1}{N}\pi_\Lambda^{ij}D_i(\ci)_{\pi j} +  \frac{1}{2N}\left( \dot{c}_3\frac{\pi_h\pi_\Lambda}{\sqrt{h}} + \dot{c}_4\frac{\pi_{hij}\pi_\Lambda^{ij}}{\sqrt{h}}\right) \ .
\end{equation}
If we add these terms to $\ca$, they do not affect the constraint (\ref{t2})
since the new terms (\ref{ca3}) are independent of $\Lambda_{ij}$.
So $\ca$ is constructed by the sum of elements (\ref{ca1}), (\ref{ca2}) and (\ref{ca3}) as shown in (\ref{ca}).

With these $\cph$ and $\ca$,  we obtain the following weak equalities:
\begin{align}
\label{}
    &\bar{\tilde{\pi}}_h^{ij}\approx \tilde{\pi}_h^{ij} = \pi_h^{ij} -\frac{\sqrt{h}}{2}F_\Lambda^{''ij}\approx 0 \ ,  \nonumber \\
    &N\sqrt{h}\left( F_\Lambda^{'ij}\Lambda_{ij} -F'\right) + \delta {\cal H}\approx N\sqrt{h}\left(  F_\Lambda^{''ij}\Lambda_{ij} -F''\right)  \ ,
\end{align}
and therefore, the transformations (\ref{H'})--(\ref{F''}) are realized.

\section{Calculations for the Horndeski/GLPV theory}

We give the detailed calculation of the explicit forms of $F''$ for the Horndeski and GLPV theories. The functional $F''$ is given in Eq. (\ref{F5}),
\begin{align}
F'' &\approx F - \epsilon \, \delta F''  \ ,\nonumber \\
     \delta F'' &= \left\{ F_NP_N +aF_h +\left( a-\frac{P_N}{N}\right)F_\Lambda^{ij}\Lambda_{ij}
    +\frac{\dot{a}}{2N}F_\Lambda -\frac{1}{N}F_\Lambda^{ij}D_i\left(P_j + (\ci)_{\pi j}\right) \right. \nonumber \\
         &\quad+ \left( F_N - \frac{1}{N}F_\Lambda^{ij}\Lambda_{ij}\right)(\cn)_{\pi_N} +c_1\Lambda + \frac{\dot{c}_1}{N} -2c_2G^{ij}\Lambda_{ij} + \frac{\dot{c}_2}{N}R \nonumber \\
         &\quad+c_3F_\Lambda\left( F_h + F_\Lambda^{ij}\Lambda_{ij}-\frac{1}{4}F_\Lambda\Lambda\right)+\frac{\dot{c}_3}{4N}F_\Lambda^2
         +\left. c_4F_\Lambda^{ij}\left( F_{hij} + {F_\Lambda}_i^k\Lambda_{kj}-\frac{1}{4}F_{\Lambda ij}\Lambda \right)+\frac{\dot{c}_4}{4N}F_\Lambda^{ij}F_{\Lambda ij}\right\} \ . \nonumber
\end{align}
The functional $F$ for the two theories is given as
\begin{equation}
\label{}
    F= A_2+A_3\Lambda+A_4\Lambda_2+A_5\Lambda_3+ B_4R+B_5G^{ij}\Lambda_{ij}  \ . \nonumber
\end{equation}
The first order derivatives of $F$ with respect to $N$, $h_{ij}$ and $\Lambda_{ij}$ are
\begin{align}
    F_N&= \frac{1}{N\sqrt{h}}\int d^3x' \frac{\delta (N\sqrt{h}F)(\vxx)}{\delta N} \nonumber  \\
    &= (A_2)_N+ \frac{A_2}{N} + \left( (A_3)_N + \frac{A_{3}}{N}\right)\Lambda
     + \left( (A_4)_N + \frac{A_{4}}{N}\right)\Lambda_2 +\left( (A_5)_N + \frac{A_{5}}{N}\right)\Lambda_3  \nonumber \\
    & + \left( (B_4)_N + \frac{B_{4}}{N}\right)R + \left( (B_5)_N + \frac{B_{5}}{N}\right)G^{ij}\Lambda_{ij}   \ , \\
    F_h^{ij}&= \frac{1}{N\sqrt{h}}\int d^3x' \frac{\delta (N\sqrt{h}F)(\vxx)}{\delta h_{ij}} \nonumber  \\
    &= \frac{1}{2}A_2h^{ij} + A_3\left( \frac{1}{2}\Lambda h^{ij} - \Lambda^{ij}\right) + A_4\left\{ \frac{1}{2}\Lambda_2h^{ij} -2\left( \Lambda\Lambda^{ij} -\Lambda^{(i}_k\Lambda^{kj)}\right)\right\}  \nonumber \\
    &  +  A_5\left\{ \frac{1}{2}\Lambda_3h^{ij} -3\left(\Lambda^2\Lambda^{ij} -[\Lambda^2]\Lambda^{ij}-2\Lambda\Lambda^{(i}_k\Lambda^{kj)}+2\Lambda_{kl}\Lambda^{(ik}\Lambda^{lj)}\right)\right\} \nonumber \\
    &- B_4G^{ij} +\frac{1}{N}\left(D^iD^j -h^{ij}D^2\right)(NB_4) + \frac{1}{2}B_5\left( G^{kl}\Lambda_{kl}h^{ij} + R\Lambda^{ij} + R^{ij}\Lambda -2R^{(ik}\Lambda_k^{j)}- 2R^{k(ilj)}\Lambda_{kl} \right) \nonumber \\
    &-\frac{1}{2N}\left\{ \left( D^iD^j -h^{ij}D^2\right)(NB_5\Lambda)+ D^2(NB_5\Lambda^{ij})
    + h^{ij}D_kD_l(NB_5\Lambda^{kl}) - 2D_kD^{(i}(NB_5\Lambda^{kj)})\right\} \ ,\\
    F_\Lambda^{ij}&= \frac{1}{N\sqrt{h}}\int d^3x' \frac{\delta (N\sqrt{h}F)(\vxx)}{\delta \Lambda_ {ij}} \nonumber  \\
    &= A_3h^{ij}
    + 2A_4\left(\Lambda h^{ij}-\Lambda^{ij}\right) +  3A_5\left(\Lambda_2h^{ij} -2\Lambda\Lambda^{ij}+ 2\Lambda^{(i}_k\Lambda^{kj)}\right)+ B_5G^{ij}\ .
\end{align}
We find that $\delta F''$ includes up to the fifth powers of $\Lambda_{ij}$.
The fifth order power terms of $\Lambda_{ij}$ in $F''$ are
\begin{align}
\label{FF5}
    \delta F''&\supset \left( F_N - \frac{1}{N}F_\Lambda^{ij}\Lambda_{ij}\right)(\cn)_{\pi_N}+ c_3F_\Lambda\left( F_h + F_\Lambda^{ij}\Lambda_{ij}-\frac{1}{4}F_\Lambda\Lambda\right) \nonumber \\
         &\qquad\quad +c_4F_\Lambda^{ij}\left( F_{hij} + {F_\Lambda}_i^k\Lambda_{kj}-\frac{1}{4}F_{\Lambda ij}\Lambda \right) \nonumber \\
       &\supset \left\{ \left( (A_5)_N +\frac{A_5}{N}\right)\Lambda_3 -\frac{3A_5}{N}\Lambda_3\right\}(\cn)_{\pi_N} + 3c_3A_5^2\Lambda_2\left( -\frac{3}{2}\Lambda_3 +3\Lambda_3 -\frac{3}{4}\Lambda\Lambda_2 \right) \nonumber \\
       &+ 3c_4A_5^2\left( \Lambda_2h^{ij}-2\Lambda\Lambda^{ij}+ 2\Lambda^{(ik}\Lambda_k^{j)}\right)\left\{\frac{1}{2}\Lambda_3h_{ij} -\frac{3}{4}\left( \Lambda\Lambda_2h_{ij} -2\Lambda^2\Lambda_{ij}+ 2\Lambda\Lambda_{il}\Lambda^l_j\right)\right\} \nonumber \\
      &\approx \left( (A_5)_N -\frac{2A_5}{N}\right)\Lambda_3 \cdot(\cn)_{\pi_N}\left[\frac{\sqrt{h}}{2}F_\Lambda^{ij}\right]
           + \frac{9}{2}c_3A_5^2\left( \Lambda_2\Lambda_3 -\frac{1}{2}\Lambda\Lambda_2^2\right) \nonumber \\
       &+ \frac{3}{4}c_4A_5^2\left\{ 2\Lambda_2\Lambda_3 + \Lambda\left( \Lambda_2^2 -4\Lambda^2[\Lambda^2] +8\Lambda[\Lambda^3] - 4[\Lambda^4]\right) \right\} \ .
\end{align}
The fourth and third power terms of $\Lambda_{ij}$ as well as $P^i$ term when $\cn=c_3=c_4=0$ are
\begin{align}
\label{FF4}
    \delta F''|_{\cn, c_3,c_4=0}&= F_NP_N +aF_h + \frac{\dot{a}}{2N}F_\Lambda+\left( a-\frac{P_N}{N}\right)F_\Lambda^{ij}\Lambda_{ij}
     -\frac{1}{N}F_\Lambda^{ij}D_i\left(P_j + (\ci)_{\pi j}\right) \nonumber \\
         & \quad+c_1\Lambda + \frac{\dot{c}_1}{N} -2c_2G^{ij}\Lambda_{ij} + \frac{\dot{c}_2}{N}R
         \nonumber \\
         & \supset -\frac{3}{N}A_5\left(\Lambda_2h^{ij} -2\Lambda\Lambda^{ij}+ 2\Lambda^{(i}_k\Lambda^{kj)}\right)D_i\left(P_j + (\ci)_{\pi j}\right) \nonumber \\
          &\quad + P_N\left( (A_5)_N + \frac{A_{5}}{N}\right)\Lambda_3 -\frac{3}{2}aA_5\Lambda_3 +3\left( a-\frac{P_N}{N}\right)A_5\Lambda_3  \nonumber \\
          &\approx -\frac{3}{N}A_5\left(\Lambda_2h^{ij} -2\Lambda\Lambda^{ij}+ 2\Lambda^{(i}_k\Lambda^{kj)}\right)D_i\left(P_j + (\ci)_{\pi j}\left[\frac{\sqrt{h}}{2}F_\Lambda^{ij}\right] \right) \nonumber \\
      &\quad+ \left\{P_N\left( (A_5)_N - \frac{2A_{5}}{N}\right) +\frac{3}{2}aA_5\right\}\Lambda_3 \ .
\end{align}
%
%

\bibliographystyle{apsrmp}

\begin{thebibliography}{99}

\bibitem{Brans:1961sx}
  C.~Brans and R.~H.~Dicke,
  Phys.\ Rev.\  {\bf 124}, 925 (1961).
  doi:10.1103/PhysRev.124.925

  \bibitem{Horndeski:1974wa}
  G.~W.~Horndeski,
  Int.\ J.\ Theor.\ Phys.\  {\bf 10}, 363 (1974).
  doi:10.1007/BF01807638

  \bibitem{Dvali:2000hr}
  G.~R.~Dvali, G.~Gabadadze and M.~Porrati,
  Phys.\ Lett.\ B {\bf 485}, 208 (2000)
  doi:10.1016/S0370-2693(00)00669-9
  [hep-th/0005016].

\bibitem{ArkaniHamed:2003uy}
  N.~Arkani-Hamed, H.~C.~Cheng, M.~A.~Luty and S.~Mukohyama,
  JHEP {\bf 0405}, 074 (2004)
  doi:10.1088/1126-6708/2004/05/074
  [hep-th/0312099].

 \bibitem{Alishahiha:2004eh}
  M.~Alishahiha, E.~Silverstein and D.~Tong,
  Phys.\ Rev.\ D {\bf 70}, 123505 (2004)
  doi:10.1103/PhysRevD.70.123505
  [hep-th/0404084].


 \bibitem{Horava:2009uw}
  P.~Horava,
  Phys.\ Rev.\ D {\bf 79}, 084008 (2009)
  doi:10.1103/PhysRevD.79.084008
  [arXiv:0901.3775 [hep-th]].

\bibitem{Blas:2010hb}
  D.~Blas, O.~Pujolas and S.~Sibiryakov,
  JHEP {\bf 1104}, 018 (2011)
  doi:10.1007/JHEP04(2011)018
  [arXiv:1007.3503 [hep-th]].

  \bibitem{Colombo:2014lta}
  M.~Colombo, A.~E.~Gumrukcuoglu and T.~P.~Sotiriou,
  Phys.\ Rev.\ D {\bf 91}, no. 4, 044021 (2015)
  doi:10.1103/PhysRevD.91.044021
  [arXiv:1410.6360 [hep-th]].



 \bibitem{Deffayet:2009wt}
  C.~Deffayet, G.~Esposito-Farese and A.~Vikman,
  Phys.\ Rev.\ D {\bf 79}, 084003 (2009)
  doi:10.1103/PhysRevD.79.084003
  [arXiv:0901.1314 [hep-th]].


 \bibitem{Germani:2010gm}
  C.~Germani and A.~Kehagias,
  Phys.\ Rev.\ Lett.\  {\bf 105}, 011302 (2010)
  doi:10.1103/PhysRevLett.105.011302
  [arXiv:1003.2635 [hep-ph]].


  \bibitem{Padilla:2010de}
  A.~Padilla, P.~M.~Saffin and S.~Y.~Zhou,
  JHEP {\bf 1012}, 031 (2010)
  doi:10.1007/JHEP12(2010)031
  [arXiv:1007.5424 [hep-th]].


  \bibitem{Deffayet:2010qz}
  C.~Deffayet, O.~Pujolas, I.~Sawicki and A.~Vikman,
  JCAP {\bf 1010}, 026 (2010)
  doi:10.1088/1475-7516/2010/10/026
  [arXiv:1008.0048 [hep-th]].

\bibitem{Kobayashi:2010cm}
  T.~Kobayashi, M.~Yamaguchi and J.~Yokoyama,
  Phys.\ Rev.\ Lett.\  {\bf 105}, 231302 (2010)
  doi:10.1103/PhysRevLett.105.231302
  [arXiv:1008.0603 [hep-th]].






\bibitem{Kobayashi:2011nu}
  T.~Kobayashi, M.~Yamaguchi and J.~Yokoyama,
  Prog.\ Theor.\ Phys.\  {\bf 126}, 511 (2011)
  doi:10.1143/PTP.126.511
  [arXiv:1105.5723 [hep-th]].

\bibitem{Naruko:2015zze}
  A.~Naruko, D.~Yoshida and S.~Mukohyama,
  arXiv:1512.06977 [gr-qc].


\bibitem{Woodard:2006nt}
  R.~P.~Woodard,
  Lect.\ Notes Phys.\  {\bf 720}, 403 (2007)
  doi: 10.1007/978-3-540-71013-4\_14
  [astro-ph/0601672].


\bibitem{Gleyzes:2014dya}
  J.~Gleyzes, D.~Langlois, F.~Piazza and F.~Vernizzi,
  Phys.\ Rev.\ Lett.\  {\bf 114}, no. 21, 211101 (2015)
  doi:10.1103/PhysRevLett.114.211101
  [arXiv:1404.6495 [hep-th]].

\bibitem{Gao:2014soa}
  X.~Gao,
  Phys.\ Rev.\ D {\bf 90}, 081501 (2014)
  doi:10.1103/PhysRevD.90.081501
  [arXiv:1406.0822 [gr-qc]].


\bibitem{Lin:2014jga}
  C.~Lin, S.~Mukohyama, R.~Namba and R.~Saitou,
  JCAP {\bf 1410}, no. 10, 071 (2014)
  doi:10.1088/1475-7516/2014/10/071
  [arXiv:1408.0670 [hep-th]].

\bibitem{Deffayet:2015qwa}
  C.~Deffayet, G.~Esposito-Farese and D.~A.~Steer,
  Phys.\ Rev.\ D {\bf 92}, 084013 (2015)
  doi:10.1103/PhysRevD.92.084013
  [arXiv:1506.01974 [gr-qc]].


\bibitem{Bekenstein:1992pj}
  J.~D.~Bekenstein,
  Phys.\ Rev.\ D {\bf 48}, 3641 (1993)
  doi:10.1103/PhysRevD.48.3641
  [gr-qc/9211017].

\bibitem{Nojiri:2003rz}
  S.~Nojiri and S.~D.~Odintsov,
  Phys.\ Lett.\ B {\bf 576}, 5 (2003)
  doi:10.1016/j.physletb.2003.09.091
  [hep-th/0307071].

\bibitem{Gleyzes:2014qga}
  J.~Gleyzes, D.~Langlois, F.~Piazza and F.~Vernizzi,
  JCAP {\bf 1502}, 018 (2015)
  doi:10.1088/1475-7516/2015/02/018
  [arXiv:1408.1952 [astro-ph.CO]].

\bibitem{Langlois:2015cwa}
  D.~Langlois and K.~Noui,
  JCAP {\bf 1602}, no. 02, 034 (2016)
  doi:10.1088/1475-7516/2016/02/034
  [arXiv:1510.06930 [gr-qc]].


\bibitem{Langlois:2015skt}
  D.~Langlois and K.~Noui,
  arXiv:1512.06820 [gr-qc].

\bibitem{Crisostomi:2016tcp}
  M.~Crisostomi, M.~Hull, K.~Koyama and G.~Tasinato,
  JCAP {\bf 1603}, no. 03, 038 (2016)
  doi:10.1088/1475-7516/2016/03/038
  [arXiv:1601.04658 [hep-th]].

  \bibitem{Crisostomi:2016czh}
  M.~Crisostomi, K.~Koyama and G.~Tasinato,
  arXiv:1602.03119 [hep-th].

\bibitem{Bettoni:2013diz}
  D.~Bettoni and S.~Liberati,
  Phys.\ Rev.\ D {\bf 88}, 084020 (2013)
  doi:10.1103/PhysRevD.88.084020
  [arXiv:1306.6724 [gr-qc]].

\bibitem{Chameleon}
  P.~Brax, C.~van de Bruck, A.~C.~Davis, J.~Khoury and A.~Weltman,
  Phys.\ Rev.\ D {\bf 70}, 123518 (2004)
  doi:10.1103/PhysRevD.70.123518
  [astro-ph/0408415].

\bibitem{Kobayashi:2014ida}
  T.~Kobayashi, Y.~Watanabe and D.~Yamauchi,
  Phys.\ Rev.\ D {\bf 91}, no. 6, 064013 (2015)
  doi:10.1103/PhysRevD.91.064013
  [arXiv:1411.4130 [gr-qc]].

\bibitem{Koyama:2015oma}
  K.~Koyama and J.~Sakstein,
  Phys.\ Rev.\ D {\bf 91}, 124066 (2015)
  doi:10.1103/PhysRevD.91.124066
  [arXiv:1502.06872 [astro-ph.CO]].

\bibitem{Saito:2015fza}
  R.~Saito, D.~Yamauchi, S.~Mizuno, J.~Gleyzes and D.~Langlois,
  JCAP {\bf 1506}, 008 (2015)
  doi:10.1088/1475-7516/2015/06/008
  [arXiv:1503.01448 [gr-qc]], and private communication with R. Saito

\bibitem{Arnowitt:1962hi}
  R.~L.~Arnowitt, S.~Deser and C.~W.~Misner,
  Gen.\ Rel.\ Grav.\  {\bf 40}, 1997 (2008)
  doi:10.1007/s10714-008-0661-1
  [gr-qc/0405109].

\bibitem{Gao:2014fra}
  X.~Gao,
  Phys.\ Rev.\ D {\bf 90}, 104033 (2014)
  doi:10.1103/PhysRevD.90.104033
  [arXiv:1409.6708 [gr-qc]].

\bibitem{Domenech:2015tca}
  G.~Dom\`enech, S.~Mukohyama, R.~Namba, A.~Naruko, R.~Saitou and Y.~Watanabe,
  Phys.\ Rev.\ D {\bf 92}, no. 8, 084027 (2015)
  doi:10.1103/PhysRevD.92.084027
  [arXiv:1507.05390 [hep-th]].

\bibitem{Mukohyama:2015gia}
  S.~Mukohyama, R.~Namba, R.~Saitou and Y.~Watanabe,
  Phys.\ Rev.\ D {\bf 92}, no. 2, 024005 (2015)
  doi:10.1103/PhysRevD.92.024005
  [arXiv:1504.07357 [hep-th]].

\bibitem{Chamseddine:2013kea}
  A.~H.~Chamseddine and V.~Mukhanov,
  JHEP {\bf 1311}, 135 (2013)
  doi:10.1007/JHEP11(2013)135
  [arXiv:1308.5410 [astro-ph.CO]].

\bibitem{Harashima}
 A.~Harashima,
 Analytical mechanics (Shokabo, 1973)


\end{thebibliography}

\end{document}